\documentclass[manuscript,authorversion,nonacm]{acmart}
\AtBeginDocument{%
  }

\setcopyright{none}
\usepackage{longtable}
\usepackage{tabularx}
\usepackage[table]{xcolor}
\usepackage{enumitem}

\newcommand{\tbf}[1]{\textbf{#1}}
\newcolumntype{Y}{>{\centering\arraybackslash}X}
\newcolumntype{Z}{>{\raggedright\arraybackslash}X}

\usepackage{ifthen}
\newcommand{\cc}[1]{
    \IfStrEq{#1}{N/A}
    {\cellcolor{gray!25}#1}
    {\ifthenelse{#1 < 50}
        {\cellcolor{red!30}#1\hspace{-0.1cm}}
        {\ifthenelse{#1 < 75}
            {\cellcolor{yellow!30}#1\hspace{-0.1cm}}
            {\cellcolor{green!30}#1\hspace{-0.1cm}}
        }}
}

\begin{document}

\title{Quality Assessment of Public Summary of Training Content for GPAI models required by AI Act Article 53(1)(d)}

\author{Dick A. H. Blankvoort}
\authornote{Both authors contributed equally to this work.}
\email{dick.blankvoort@ru.nl}
\orcid{0009-0003-0766-4678}

\affiliation{%
  \institution{AI Accountability Lab (AIAL), ADAPT Centre, Trinity College Dublin}
  \city{Dublin}
  \country{Ireland}
}

\author{Harshvardhan J. Pandit} 
\authornotemark[1]
\email{me@harshp.com}
\orcid{0000-0002-5068-3714}

\affiliation{%
  \institution{AI Accountability Lab (AIAL), ADAPT Centre, Trinity College Dublin}
  \city{Dublin}
  \country{Ireland}
}

\author{Maximilian Gahntz}
\email{maximilian@mozillafoundation.org}
\affiliation{%
  \institution{Mozilla.org}
  \state{California}
  \country{USA}
}

\renewcommand{\shortauthors}{Blankvoort \& Pandit \& Gahntz}

\newcolumntype{P}[1]{>{\centering\arraybackslash}p{#1}}

\begin{abstract}
  The AI Act's Article 53(1)(d) requires providers of general-purpose AI (GPAI) models to publish a sufficiently detailed public summary about the content used for training based on a template provided by the AI Office. The stated goal of this obligation is to increase transparency regarding the data used for training GPAI models, and to enable relevant stakeholders to exercise their rights, especially regarding IP, copyright, and data protection. This paper provides a quality assessment framework to assess the public summary across two key dimensions: \textit{transparency} regarding information being provided in a clear, comprehensive, and sufficiently detailed manner; and \textit{usefulness} regarding whether the provision of the document and the contents can be effectively utilised by stakeholders to carry out rights related actions. This framework enables identification of key issues in public summaries, and provides a structured and research-based method to compare practices across public summaries and providers. It also enables authorities such as the AI Office to identify potential issues that could emerge and provides actionable recommendations and guidelines for providers to develop public summaries with high quality. The paper provides an assessment of 5 public summaries published as of 12th January 2026 which were found through an exhaustive search process. To disseminate these findings as a public resource, the paper also describes the development of a website where the assessments, outcomes, and methodologies will be shared.
\end{abstract}


\maketitle

\section{Introduction}
The EU's Artificial Intelligence Act (AI Act)~\cite{ai_act} is the world's first regulation that specifically targets AI technologies. Drafted over a period of 3 years, the regulation saw drastic changes due to rapid advancements made in the field of General Purpose AI (GPAI) – in particular the launch of powerful large language models (LLMs) and their risk to rights. The AI Act itself mentions the potential for GPAI models to pose `systemic risks' to public health and safety, democratic processes, and economic security (Recital 110). In addition to these, it also mentions the challenges presented by GPAI models to artists and other stakeholders due to the inclusion of intellectual property and copyrighted material in training data (Recital 105). It also notes the responsibilities of GPAI providers and requires transparency measures, including those that involve the public release of material (Recital 107).

While most requirements are expected to be complied with through internal and confidential mechanisms and documentations, one of the publicly accessible obligations includes the requirement to publish ``\textit{a sufficiently detailed summary about the content used for training of the general-purpose AI model}'' (Article 53(1)(d)). The justification for this, as noted in Recital 107, is to increase transparency regarding the data used in the pre-training and training phases of model development, and to facilitate rights enforcement associated with IP, copyright, and other laws. The European Artificial Intelligence Office (AI Office), established within the European Commission for implementation of the AI Act, is responsible for providing a common template to allow implementation of the obligation.

The public summary of training content is an important document for stakeholders, as emphasised in the explanatory notice, because of its role in ensuring accountability of GPAI providers and in enabling exercise of fundamental rights investigations and enforcement actions under EU laws. Therefore, it is important for the document to be of a sufficiently high quality to enable these intended investigations and accountability measures. The AI Office published the Template for the Public Summary of Training Content for general-purpose AI models on July 24th, 2025~\cite{template_presentation}, along with an explanatory notice that described the objectives of the obligation, the role of the provided template, and the considerations regarding provider's need for confidentiality being balanced with the interests of stakeholders. 
However, neither the AI Office's explanatory note nor its guidelines~\cite{guidelines_aiact} provide an indication of how stakeholders can assess and ensure that the public summary is sufficiently transparent and useful. 
In addition, despite the template being published over six months prior at the time of writing, and many major GPAI providers having published updated models, none have yet published their public summaries~\cite{openfuture_notransparency}. 
Thus, achieving the desired transparency, accountability, and rights empowerments through the public summary risk are at risk for avoision tactics \cite{yew_red_2025}, including bad faith implementations, intentional obfuscations, and delaying investigations and enforcements. 

The public summary obligation has also received little attention in academic research related to the AI Act, with no studies or analysis available that can guide how the public summaries should be published or how to assess adherence to the AI Act's principles. Providers thus also may not have sufficient understanding of how to ensure their public summaries are of a sufficient quality and can achieve the intended outcomes. Simultaneously, the AI Office might prioritise (other) urgent obligations related to prohibited and high-risk AI systems and the integration of harmonised standards, and not consider public summaries a key focus in terms of assigning resources or proactive investigations. An understanding of how the published summaries should be filled so as to fulfil the intended objectives outlined in the AI Act would thus benefit both the Providers as well as enforcers like the AI Office.

To assist with the upcoming enforcement of AI Act and ensure the intended goals of requiring public transparency and accountability are achieved, we provide a `documentation quality assessment framework'. This framework allows Providers to assess their implementations against a comprehensive set of metrics and ensure the document is of high quality. It also allows auditors, evaluators, and enforcers -- including the AI Office -- to objectively evaluate published summaries in a structured and comprehensive manner to identify deficiencies. Additionally, the framework also serves as a tool to compare published summaries across models and providers, and in doing so to identify specific trends that can guide future revisions to the template made by the AI Office.
In summary, we provide three core contributions:

\begin{enumerate}[label=RC\arabic*]
    \item A Documentation Quality Assessment Framework for evaluating the public summaries of training content as required by AI Act Article 53(1)(d) for achieving transparency, accountability, and rights empowerment;
    \item An evaluation of five published summaries using the developed framework that shows which aspects of the template are filled in sufficiently while also exposing which parts providers face difficulties in;
    \item A discussion on the development and application of our framework that provides recommendations for Providers to improve their documentation practices and suggestions for the AI Office for improving future templates.
\end{enumerate}

The rest of the paper is structured as follows: Section 2 presents background information regarding the AI Act Article 53(1)(d) and template published by the AI Office, and a literature review on documentation quality assessments. Section 3 presents the methodology and the completed documentation quality assessment framework. Section 4 presents findings from applying the framework to published public summaries. Section 5 discusses our findings and providers recommendations for Providers and the AI Office based on our experiences in developing and utilising the framework. Section 6 concludes the paper.

\section{Background \& Related Work}

\subsection{AI Act Article 53(1)(d) and Template for Public Summary}

The AI Act's Article 53(1)(d) puts an obligation on General Purpose AI (GPAI) model providers (henceforth just `Providers') to publish a summary of training content used to develop the model (henceforth just `summary'). The obligation also requires the summary to based on the common template provided by the AI Office.
The rationale for this, as stated in Recital 107, is `to facilitate parties with legitimate interests ... to exercise and enforce their rights', and for which the provided summary must be `generally comprehensive in its scope instead of technically detailed'. 
As with the rest of the AI Act, the obligation to publish the public summary of training content will be applicable in a staggered manner~\cite{ai_act} from 2nd August 2025.
However, the AI Office will start enforcement actions only from 2nd August 2026, providing a one year leeway for providers to implement the obligation.
For GPAI models `placed on the market' prior to this date, the summaries must be made available before 2nd August 2027. 
In practical terms, providers who publish new models must also publish the public summary of training content in accordance with the AI Act, or face enforcement actions and potential penalties once enforcement begins.
In addition, the AI Act requires the summaries to be made available in the `same context' as the model, which in practical terms would mean in the same website or location, in a way that would be reasonable for stakeholders to discover.


An explanatory notice and template for the public summary of training content for GPAI models required by Article 53(1)(d) of the AI Act was published on 24th July 2025~\cite{template_presentation}. The announcement stated it to be a complement to the guidelines on the scope of the rules for general-purpose AI models published earlier on 18th July 2025~\cite{template_presentation} and the General-Purpose AI Code of Practice released on 10th July 2025~\cite{code_of_practice}. The template forms a necessary part of the obligation to implement Article 53(1)(d), while the guideline and code of conduct are intended as voluntary measures by Articles 95 and 96. The explanatory note accompanying the public summary template contains 34 paragraphs that outline the objective of the template, its development process, scope of details for filling in the template, balancing of business confidentiality, guidelines for modifications and updates to models, and procedural aspects for the implementation of the AI Act. Paragraph 34 states the Commission (through the AI Office) will monitor the implementation of the template and has the option of conducting a review prior to the start of enforcement actions on August 2nd 2026 based on technical developments and practical experiences.

The explanatory note also provides insight into how the template was drafted (§5) and its intended objectives that guided the final design (§15,§16).  
The public summary template (henceforth just ``template") is a 6-page Office Open XML Document (\textit{.docx} file used in word processing applications such as Microsoft Word and LibreOffice Writer). It contains three sections: Section 1 (General Information) asks about provider identification, model details, and broad information about training data modalities; Section 2 (List of Data Sources) contains 6 subsections that enquire about data sources based on their characteristics as public or non-public sources, data crawled and scraped from online sources, reuse of user data, and synthetic data; and Section 3 (Data processing aspects) asks about data processing aspects relevant to text and data mining limitations and removal of illegal content. 

\subsection{AI \& Technical Documentation Quality}
The assessment of the summary requires an approach that considers the role of the summary as documentation of the GPAI model alongside its utility as a notice for relevant stakeholders to enforce their rights and protect their legitimate interests in accordance with the AI Act. Therefore in addition to assessing the `quality' of information within the summary, we must also assess the manner in which the summary is provided and assess the broader `document quality'. 
For this, we searched for existing approaches that provided guidance or assessment resources related to the AI Act. We found efforts such as those by HuggingFace~\cite{model_card_regulatory_tech} to assess model cards in a repository based on section titles that reflected industry best practices in terms of maintaining and assessing documentation. \citet{liangSystematicAnalysis321112024} provided a comprehensive and systematic analysis of 32,111 AI model documentation pages on HuggingFace and found only 44.2\% models had a corresponding model card, and that there is an overwhelming lack of conformity regarding structure, contents, and expected information as compared to established guidelines. The paper also discusses resources including tools to generate model cards, guidelines for filling in model cards, and educational materials. \citet{lucajTechOpsTechnicalDocumentation2025} provided templates for AI Act documentation, which draw from the experience of dealing with model cards, but only focus on inter-organisational documentation. From this literature we draw the conclusion that current industry practices do not include good quality documentation, especially those that are to be made publicly available, and that this represents a risk regarding the availability and the utility of public summaries to be published by the same providers.
Of note, we did not find any work specifically related to Article 53(1)(d) public summaries. 

Therefore, we looked towards other broader approaches regarding quality assessment associated with technical documentation, in particular those associated with software. To enhance our understanding on both topics, we utilised authoritative sources of well-established practices such as the ISO Quality Management Systems (QMS)~\cite{iso_9001} and ISO/IEC Systems and software Quality Requirements and Evaluation (SQuaRE)~\cite{isoiec_25052} standards, as well as the Eurostat Handbook on Data Quality Assessment Methods and Tools~\cite{bergdahl2007handbook}. 
A simplified explanation of how quality assessment frameworks are developed is to identify a particular set of `dimensions' (similar to principles) that best represent the quality in context of the objectives of the assessments, and to then identify specific `metrics' (e.g. questions) that should be evaluated to score each dimension.
We therefore conducted a literature review of technical/software documentation quality assessments and data quality assessments for identifying established dimensions and their utility based on empirical results and user studies. 

Surveying papers from 2010 to 2025, we identified 54 papers in total, of which we chose 7 papers as being relevant. \citet{garousiEvaluatingUsageQuality2013} described the application of Action Research (AR) methodology within an industrial setting to assess maintenance and application of documented information by using the Goal-Question-Metric paradigm. \citet{dingKnowledgebasedApproachesSoftware2014} presented a comprehensive literature review of 60 articles to derive 12 quality attributes, four cost categories, and nine benefit categories. \citet{ploschValueSoftwareDocumentation2014} conducted a survey with 88 experts and identified the most important quality attributes for documentation as accuracy (as a synonym for correctness), clarity, consistency, readability, structuredness, and understandability. \citet{treudeAccuracyAssessingSoftware2020} conducted a similar study where they asked technical editors to rate quality metrics for different types of documentation. This study provides important insights for the assessment of public summaries as narrative documentation with technical information. \citet{aghajaniSoftwareDocumentationIssues2019} analysed 878 documentation artefacts and developed a taxonomy of (common) documentation issues that was mapped to specific quality attributes. \citet{tangEvaluatingSoftwareDocumentation2023} presented a framework of 9 metrics developed through design co-creation process with developers and use this to automate the assessment and verification of software documentation. \citet{zaveriQualityAssessmentLinked2015} presented a comprehensive survey of linked data quality assessment, which provided specific metrics and dimensions for assessment of the public summary as a document linked to the model as well as the use of hyperlinked information within it. We also considered the FAIR Principles~\cite{wilkinsonFAIRGuidingPrinciples2016} which are considered best practice within the broader scientific community for the open sharing of information.

\begin{table}
    \centering
    \caption{Overview of dimensions identified from survey of literature related to technical documentation, software documentation, and data quality assessment frameworks}
    \begin{tabularx}{\textwidth}{|c|Y|Y|Y|Y|Y|Y|Y|}
    \hline
    \tbf{Dimension} & \tbf{Garousi et al} & \tbf{Ding et al} & \tbf{Plosch et al} & \tbf{Treude et al} & \tbf{Aghajani et al} & \tbf{Tang and Nedi} & \tbf{Zaveri et al} \\\hline
    \tbf{Accessibility} & & x & & & x & x & x \\\hline
    \tbf{Accuracy} & x & x & x & x & & & x \\\hline
    \tbf{Availability} & & & & & x & & x \\\hline
    \tbf{Clarity} & & x & x & x & x & & \\\hline
    \tbf{Conciseness} & & x & x & x & & & x \\\hline
    \tbf{Completeness} & x & x & x & & x & x & x \\\hline
    \tbf{Consistency} & x & x & x & x & x & x & x \\\hline
    \tbf{Correctness} & & x & x & & x & x & x \\\hline
    \tbf{Readability} & x & & x & & x & x & x \\\hline
    \tbf{Relevancy} & x & & & & & x & x \\\hline
    \tbf{Structure} & x & & x & x & x & x & \\\hline
    \tbf{Traceability} & & x & x & & x & x & \\\hline
    \tbf{Understandability} & & x & x & x & & x & x \\\hline
    \tbf{Up to date-ness} & & x & & & x & x & x \\\hline
    \tbf{Usability} & & x & & & x & x & x \\\hline
    \tbf{Usefulness} & & x & & & x & x & x \\\hline
    \end{tabularx}
    \label{tab:metrics-lit}
\end{table}

Table 1 summarises our findings regarding demonstrably useful dimensions for assessing and representing quality of technical documentation, software documentation, and data(sets). For convenience, we consolidated dimensions based on their provided definitions, e.g. we combined comprehension and understandability based on equivalence in their definitions. 
We only included metrics that were used in at least 3 frameworks to ensure consistency across frameworks. Doing so resulted in 16 metrics that represent useful and validated indicators for quality of documentation, and which are suitable for use in assessing the public summary. 

\section{Documentation Quality Assessment Framework}
\subsection{Development Methodology}
To develop our quality assessment framework, we first assessed the contents and structure of the template by discussing the explanatory note and the structure and contents of the template. We then established an understanding of what information would be filled into the template to create the public summary based on our existing experiences in analysis of GPAI models and their documentation as well as understanding of the goals and requirements stated in the AI Act. 
For practical reasons, we identified two use-cases that guided our design process: `good faith' implementations where all fields are filled in and it is clear from the information that the providers have made a conscientious effort; and `bad faith' implementations where not all fields may be filled in or where there is unclear, ambiguous, or otherwise unhelpful information in the public summary. Our objective is to support the `good faith' public summaries in improving their quality while highlighting deficiencies present in the `bad faith' implementations.

Following an iterative development process, we first identified the high-level categories through which the quality of the public summary should be represented. Based on the rationale provided by the European Commission in its explanatory note as well as the role of the summary as defined in the AI Act, we identified \textit{Transparency} and \textit{Usefulness} as good overall indicators for intended stakeholders to understand the quality of the public summary. In this, \textit{transparency} refers to the extent of information provided through the public summary, and \textit{usefulness} represents the utility of provided information as well as its potential to be actionable by specific stakeholders. We selected \textit{transparency} due to its existing and established application in the context of legal documentation, for example as a principle in the GDPR, as well as being listed as an objective of methods such as model cards. And we chose \textit{usefulness}, which is well-defined as a quality dimension in literature, as a broad measure of whether the provided information is sufficient to take meaningful actions and ensure accountability. Using our framework, a \textit{`high-quality'} public summary is described as a document that provides information in a clear, structured, accurate, and consistent manner (\textit{transparency}) such that it allows relevant stakeholders to understand the training process and to take relevant actions where necessary (\textit{usefulness}). Public summaries that do not meet these requirements are considered as \textit{`low-quality'}, and are potentially in breach of the template and the corresponding Article 53(1)(d) obligations.

\begin{table}
    \centering
    \caption{Chosen quality dimensions and their categorisation under \textit{Transparency} and \textit{Usefulness} as overall quality indicators.}
    \begin{tabularx}{\textwidth}{|p{1.9cm}|p{1.70cm}|Z|}
    \hline
    \textbf{Dimension} & \textbf{Group} & \textbf{Description} \\\hline
    Clarity & Transparency & information is unambiguous, easy to understand, and avoids misinterpretation \\\hline
    Completeness & Transparency & relevant and necessary information is provided \\\hline
    Consistency & Transparency & terminology, format, and structure is consistent within and across documents \\\hline
    Correctness & Transparency & information is accurate, validated, and reliable \\\hline
    Accessibility & Usefulness & information is easy to obtain, navigate, interact, \& includes accessibility standards \\\hline
    Comprehension & Usefulness & information is understandable and interpretable for the intended audience \\\hline
    \end{tabularx}
    \label{tab:dimensions}
\end{table}

To quantify how transparency and usefulness should be assessed for public summaries, we deliberated on each dimension identified in literature regarding its applicability and interpretation for the public summary. This was an iterative process where we also discussed whether the chosen dimensions would be clear or confusing for various audiences, and how they would be useful to indicate transparency and usefulness. The outcome of this was the selection of 6 metrics: \textit{Clarity}, \textit{Completeness}, \textit{Consistency}, and \textit{Correctness} for representing \textit{Transparency}, and \textit{Accessibility} and \textit{Comprehension} for representing \textit{Usefulness}. 

In addition to representing the quality of the public summary through these dimensions, we also identified the need to consider the quality dimensions of specific sections of the template. For example, relevant stakeholders may be more interested in Section 2.4 \textit{User data} as compared to 2.5 \textit{Synthetic data}, and thus would benefit from an dedicated quality score for how \textit{transparent} and \textit{useful} specific sections of the summary are for their particular needs. To implement this, we divided the template into 8 sections (see table \ref{tab:dimensions}) and assessed each section independently using the chosen dimensions. Of these, 7 sections are as they appear in the template: \textit{General information} (Section 1), \textit{Public Data Sources} (Section 2.1), \textit{Private Data Sources} (Section 2.2), \textit{Scraped/Crawled Data} (Section 2.3), \textit{User Data} (Section 2.4), \textit{Synthetic \& Other Data} (Section 2.5 \& 2.6), and \textit{Data Processing} (Section 3). In addition to these, we also added \textit{Document} as a section to represent assessments regarding metadata, provision, accessibility, and adherence of template structure of the public summary document. 

To evaluate the public summary using our chosen dimensions, we utilised the  Goal-Question-Metric paradigm (GQM)~\cite{caldiera1994goal}, which was also used in some of the analysed quality assessment frameworks from literature. In GQM, we start with high-level goals, which in our case are assessing the 6 quality dimensions for each of the 8 sections, for a total of 42 goals. For each goal, we then identify specific \textit{metrics} (i.e. questions or requirements) that are assessed and scored, and where each metric represents an independent assessment of a piece of information required by the template, and is associated with exactly one dimension. For example, though Section 1.1 has a field asking about provider name and contact, we developed two metrics to check whether name and contact are filled in independently as part of the completeness dimension. For provenance of how each metric was developed, we created and attached identifiers for each information element in the template. For example, we assigned the identifier `\textit{1.1.a}' to the provider name and contact information present in the template, and then further split these into name (`\textit{F1.1.a.1}') and contact (`\textit{F1.1.a.2}'). Using this method, we created a comprehensive set of 242 metrics for a thorough analysis of any public summary based on the template. 

\begin{table}
    \centering
    \caption{Sections; consolidated metrics; total metrics}
    \begin{tabular}{|l|l|c|}
    \hline
    \tbf{Section in our QA framework} & \tbf{Public Summary Template Sections} & \tbf{No. of metrics} \\\hline
    Document & - & 30 \\\hline
    General information & Section 1 & 54 \\\hline
    Public Data Sources & Section 2.1 & 26 \\\hline
    Private Data Sources & Section 2.2 & 27 \\\hline
    Scraped/Crawled Data & Section 2.3 & 46 \\\hline
    User Data & Section 2.4 & 14 \\\hline
    Synthetic \& Other Data & Section 2.5 \& 2.6 & 28 \\\hline
    Data Processing & Section 3 & 17 \\\hline
     & \textit{Total} & \textit{242} \\\hline
    \end{tabular}
    \label{tab:sections-metrics}
\end{table}

\subsection{Assessment Procedure}

To assess the quality of a given public summary, the evaluator first scores each metric individually. To do this, the evaluator assesses the corresponding information field and element, and records a value that reflects their findings as: \textit{sufficient} (scored as \textit{1}) to indicate the provided information meets or satisfies the requirement e.g. by being clear, or being complete; \textit{insufficient} (scored as \textit{0}) to indicate either an absence of information or where the information does not meet the criteria of the metric e.g. by being obfuscated or being incomprehensible; and \textit{partially sufficient} (scored as \textit{0.5}) to indicate that information partially satisfies the requirement. As not all questions in the template carry the same importance or impact, we assigned an `\textit{importance score}' for each question (called `\textit{weight}' in quality assessment literature). For example, key questions in the data sources section that ask for specific details have a higher score due to their perceived importance as compared to generic and optional information. The final score for each metric is the assigned score (\textit{0...1}) multiplied by the weight. By utilizing weighted scores in this manner, the quality framework proportionately rates public summaries that provide key information as being of better quality while penalizing those that don't as lower quality. Once all metrics are scored as above, each dimension within a section is scored by taking the average of all metrics associated with that dimension and section. The transparency and usefulness scores are then calculated for each section by taking the average of the scores of their respective dimensions. The overall transparency and usefulness score for the public summary are then calculated as the average of the transparency and usefulness scores for each section. Finally, a letter grade system from \textit{A} as the highest quality and \textit{F} as the lowest quality is assigned based on the scores to enable stakeholders to understand quality implications at a glance. 

The template contains optional sections which we counted as mandatory to reward those providers who do provide additional information in their summaries. Without this approach, a summary that does not provide additional information would score equally to a summary that does provide additional information. Whereas by counting optional information we incentivise providers to go beyond the minimum information requirements and provide additional details. The template also contains fields that are conditional, for example, Section 2.5 has fields that should only be filled in if the first question asks whether synthetic data was used for training the model. If the answer is yes, the rest of the fields are necessary, and should be assessed. However, if the answer is no, then not filling in these fields should not be penalised. We took this into account by only counting those metrics that are applicable during the scoring of dimensions, so that a summary where some sections do not apply is not penalised in comparison to other summaries where the sections are required. Given that this means different public summaries may have different sections filled in, we normalise the scores as a percentage to enable comparing them along the same dimensions. This methodology is fairly similar to the one used by the European Data Portal where a variety of datasets are comparable through normalised scores \cite{mqa_methodology}.

\section{Quality Assessment of Published Summaries}
To assess the utility of our framework, we sought to apply it in practice to published summaries. The AI Act only requires the public summaries to be published `in the same context' as the model, which does not lead to a specific or consistent location for where to find them. We therefore utilised a combination of search engines (using queries such as ``public summary of training content" and ``Article 53(1)(d) public summary'') and utilising experimental web-search features of GenAI services such as Google Search and ChatGPT. 
We also undertook a manual discovery process where we looked at: (1) The webpage or README in the repository where the model weights or model card are provided; (2) The legal or compliance page on the model provider's webpage; (3) The technical report or paper where the model's architecture and functioning are described. We expected the public summary to be explicitly indicated as such (i.e. titled 'public summary') or by referring to AI Act's Article 53(1)(d) obligations or broadly referring to AI Act compliance.
We first prioritised Anthropic, Google, OpenAI, Meta, Mistral, and xAI as high-profile GPAI model providers based on popularity, reach, and as signatories of the Code of Practice\footnote{We note that xAI is only signatory to the Security chapter of the Code of Practice, and has deliberately opted to not sign the Transparency chapter which is relevant to public summaries.}. We did not find any public summaries published by these providers based on the method described earlier. We then utilised the European Open Source AI Index~\cite{liesenfeld_european_2025} to identify additional models and providers. 

In total, we identified four public summaries as of 20th January 2026 – HuggingFace for their SmolLM3-3B model, Swiss National AI Initiative for their Apertus model family, Speakleash for the Bielik v3 11B Instruct model, and Bria AI for the Bria 3.2 model. We also found a document titled ``data information'' in the repo of Microsoft's Phi-4 model on HuggingFace whose structure was similar to the template, but which was not explicitly mentioned as a public summary. Though there was no indication as to this being a public summary under Article 53(1)(d), we opted to assess it due to the similarity in structure of the template, as well as with the understanding that this would be a valid expectation from stakeholders looking for the public summary within the model's repository. Euractiv's reporting in January confirms the lack of public summaries, and also includes responses from Google and OpenAI that they are working on understanding the requirements and that Mistral did not respond \cite{euractivAICompaniesAre2026}. We thus could analyse only the found published summaries.

For each identified public summary, we first created a locally archived copy (PDF) for the specific version we annotated in case they were updated. We then read through the public summary to familiarise ourselves with its structure, content, and implications. One evaluator then scored the public summary using developed metrics and recorded the outcomes in a spreadsheet. The other evaluator then worked with the first annotator to verify the assessment and make clarifications where necessary. We opted for this approach as the public summary is a new form of documentation and we had no examples when developing the initial framework, and only found the summaries closer to completion stages of the work. In this process, there was no disagreement recorded. To ensure consistency, the same persons followed the roles of evaluator and verifier for each public summary. We then calculated the scores for each section and the overall transparency and usefulness scores through the method described earlier. As these were the first instances of public summary assessment, we then compared the scores and the contents of the public summaries to understand whether our scores were fair and an appropriate indicator of what was provided in the public summaries. We did not find any pertinent issues and assessed the scores to be a fair approximation of our own subjective impressions. The scores for the five identified models are described in Table \ref{tab:finalscores}, with the full assessment provided in Appendix \ref{tab:scores-expanded}. Below we discuss the contents, our scoring, and implications for each public summary.

\begin{table}
    \centering
    \caption{Scores for published public summaries of training content. Percentages represent normalised scores, \textit{N/A} represents the dimension is not applicable and therefore was not scored. Colours are used to draw attention to severity of scores with green indicating a higher quality, orange indicating moderate quality, and red indicating low quality.}
    \begin{tabular}{|l|l|r|r|r|r|r|}
    \hline
    \tbf{Section} & \tbf{Dimension} & \tbf{SmolLM} & \tbf{Apertus} & \tbf{Bielik} & \tbf{Phi} & \tbf{Bria} \\\hline
    Document & Transparency & \cc{68}.97\% & \cc{87}.04\% & \cc{88}.89\% & \cc{87}.04\% & \cc{70}.37\%\\\hline
     & Usefulness & \cc{58}.00\% & \cc{84}.21\% & \cc{84}.21\% & \cc{81}.58\% & \cc{68}.42\% \\\hline
    General Information & Transparency & \cc{73}.53\% & \cc{74}.62\% & \cc{87}.03\% & \cc{70}.59\% & \cc{69}.73\% \\\hline
     & Usefulness & \cc{90}.91\% & \cc{93}.75\% & \cc{93}.75\% & \cc{100}.00\% & \cc{92}.86\%\\\hline
    Public Data Sources & Transparency & \cc{96}.06\% & \cc{100}.00\% & \cc{86}.91\% & \cc{3}.70\% & \cc{100}.00\% \\\hline
     & Usefulness & \cc{100}.00\% & \cc{100}.00\% & \cc{100}.00\% & \cc{0}.00\% & \cc{N/A} \\\hline
    Private Data Sources & Transparency & \cc{100}.00\% & \cc{100}.00\% & \cc{100}.00\% & \cc{53}.85\% & \cc{100}.00\% \\\hline
     & Usefulness & \cc{N/A} & \cc{N/A} & \cc{N/A} & \cc{N/A} & \cc{100}.00\% \\\hline
    Scraped/Crawled Data & Transparency & \cc{100}.00\% & \cc{100}.00\% & \cc{83}.78\% & \cc{0}.00\% & \cc{100}.00\% \\\hline
     & Usefulness & \cc{N/A} & \cc{N/A} & \cc{51}.46\% & \cc{0}.00\% & \cc{100}.00\%\\\hline
    User Data & Transparency & \cc{100}.00\% & \cc{100}.00\% & \cc{100}.00\% & \cc{0}.00\% & \cc{100}.00\% \\\hline
     & Usefulness & \cc{N/A} & \cc{N/A} & \cc{N/A} & \cc{N/A} & \cc{N/A}\\\hline
    Synthetic \& Other Data & Transparency & \cc{60}.00\% & \cc{100}.00\% & \cc{100}.00\% & \cc{8}.89\% & \cc{100}.00\% \\\hline
     & Usefulness & \cc{0}.00\% & \cc{N/A} & \cc{100}.00\% & \cc{0}.00\% & \cc{N/A} \\\hline
    Data Processing & Transparency & \cc{92}.00\% & \cc{100}.00\% & \cc{98}.72\% & \cc{52}.00\%  & \cc{100}.00\% \\\hline
     & Usefulness & \cc{100}.00\% & \cc{100}.00\% & \cc{71}.43\% & \cc{0}.00\% & \cc{100}.00\% \\\hline
    Overall Scores & Transparency & \cc{82}.50\% & \cc{92}.90\% & \cc{88}.02\% & \cc{33}.30\% & \cc{86}.97\% \\\hline
     & Usefulness & \cc{86}.01\% & \cc{97}.14\% & \cc{71}.11\% & \cc{24}.54\% & \cc{94}.74\% \\\hline
    \textit{Overall Grades} & \textit{Transparency} & \tbf{\textit{B+}} & \tbf{\textit{A\phantom{+}}} & \tbf{\textit{B+}} & \tbf{\textit{D}} & \tbf{\textit{B+}} \\\hline
     & \textit{Usefulness} & \tbf{\textit{B+}} & \tbf{\textit{A+}} & \tbf{\textit{C+}} & \tbf{\textit{F}} & \tbf{\textit{A\phantom{+}}}\\\hline
    \end{tabular}
    \label{tab:finalscores}
\end{table}

\subsection{SmolLM by HuggingFace}
The public summary for HuggingFace's SmolLM3-3B model is available on its platform as a webpage~\cite{SmolLM33BPublicSummary2025}. In general, the structure of the template was preserved with clear indication of section numbers, consistent field labels, and explicitly indicating yes/no where applicable. Other than issues with accessibility, we found no major issues with the public summary. It scored 82.50\% and received a grade of B+ for transparency, and 86.01\% with grade B+ for usefulness, which meant that the summary was of a moderate but good quality. SmolLM scored particularly well on its disclosure of public data sources, a key disclosure demanded by Article 53(1)(d), and a good indicator of transparency in model development process.

The SmolLM model described in the summary was modified from a previous model whose details were not provided in this summary. We assumed this was because of the previous model being published before the cut-off date for publishing summaries, but were uncertain regarding the correct legal interpretation for modified models where the published summaries for base model are not available. SmolLM's summary was not penalised in this regard.
One major issue we found that could be easily rectified was regarding the Section 2.5 Synthetic data, where an entire field was missing (GPAI models available on market) and its information was instead put in the next field (GPAI models not available on market). This resulted in deduction of points regarding structure and for the incorrectly filled fields. 
Another issue we faced was regarding accessibility, where despite the summary following the structure of the template correctly, accessing and exporting a copy was made particularly difficult due to the `dynamic' nature of the webpage.
This resulted in a lower document accessibility score.


\subsection{Apertus by Swiss AI Initiative}
The public summary provided by the Swiss AI Initiative covers multiple models -- Apertus-8B, Apertus-8B-Instruct, Apertus-70B, and Apertus-70B-Instruct from the Apertus model family~\cite{Apertus_EU_Public_SummarypdfSwissaiApertus70B25092025}. Like SmolLM, the public summary was also published on HuggingFace. However, instead of a dynamic webpage it was provided as a PDF closely following the template. Each field was filled in, including explicitly marking sections as not applicable, though some fields included superfluous information not relevant to the topic or question which caused a few points deduction. We assessed its score to be 92.90\% with Grade A for transparency, and 97.14\% with Grade A+ for usefulness, which were the highest of all assessed summaries.


\subsection{Bielik by Speakleash}
The open collaboration project SpeakLeash published a public summary for its Bielik v3 11B Instruct model~\cite{bielik} on its website and also linked it in its model card. The public summary follows the structure of the template, and describes Bielik as a fine-tuned version of Mistral. Of note, Bielik utilised the optional additional comments field in Section 1.3 to describe the purposes, regional, and linguistic characteristics of the dataset used as based on Polish and EU legal administrative domains and the inclusion of Silesian and Kashubian language Wikipedia pages.
For their list of public data sources, however, the description referenced external documents (`preprints') which resulted in a deduction of points as relevant information was not provided in the document. We assessed its score to be 88.02\% with grade B+ for transparency, and 71.11\% with grade C+ for usefulness.

\subsection{Phi by Microsoft}
We did not find an explicitly published summary for Microsoft's Phi-4 model, but found a file in its repository on HuggingFace called ``data\_summary\_card.md'', whose structure closely matched that of the template~\cite{Data_summary_cardmdMicrosoftPhi42025}. 
The title used in the file was ``Data Summary'', which was also title used by HuggingFace in its SmolLM public summary. 
We are uncertain as to whether this represents an industry practice of co-opting of the formal title of ``public summary of training content" present in the template with a broader and vague title that could be confused with other data-related documentation also provided with GPAI models. 
We opted to assess this model regardless based on the assumption of stakeholders viewing the document as fulfilling the template requirements, as explained earlier. 

The assessment showed that fields provided in the document are filled in to some extent, but that the document itself is quite sparse and does not provide many details and also had sections missing. We further found that section numbers and questions had a significant mismatch from what was provided in the template. The document also suffered from subjective non-relevant statements, such as question 2.3.1 on whether public data was used to train the model being answered with ``\textit{Microsoft follows all relevant laws and regulations pertaining to personal information}" – which is neither relevant nor clear. We also found the document did not provide required information significantly, for example subsequent parts that enquire about the source and uses of personal data (2.4 User data in template) were found completely missing. Our assessment reflects these systematic major issues in the outcomes where the document scored 33.30\% with a Grade D for transparency, and 24.54\% with a Grade F for usefulness, which is the lowest score amongst all assessed summaries.

\subsection{Bria by Bria AI}
The public summary for the Bria 3.2 model was published by Bria AI on its website under the `Security \& Compliance' section as a link to a PDF hosted on Google Drive~\cite{bria}.
The summary closely followed the structure of the template, but which provided contact details for the authorised representative but not for the company itself. The document also lacked some metadata such as version information and links to documentation.
The summary describes training only involving licensed data, which requires relatively little disclosure through the template. We assessed its score as 86.97\% with a grade of B+ for transparency, and 94.74\% with a grade of A for usefulness.

\section{Discussion}

\subsection{Status of Published Summaries}
Of the five summaries we assessed, three ---SmolLM, Apertus, and Bria--- had a sufficiently high quality to be considered as having made a good effort.
All three had relatively minor issues that we felt did not impact the objectives of the template, and which should be easily fixable.
Of these, Apertus scored the highest, and the evaluators consider it as an exemplary example for guiding subsequent public summaries to be published by other providers.
Of the remaining two -- Bielik and Phi -- only Bielik's public summary was explicitly labelled as such. While Bieliks's summary also seemed to be in good faith, it did sufficiently provide details in required forms and as a result suffered from a lower scores. We consider this to also be fixable - though it would require the Provider to be willing to provide this information. In the case of Phi, the document was unclear and shows several systemic issues which would require major fixes. We are therefore unclear as to whether Microsoft intended this document to be a public summary, and hope that a revised and improved version would be provided soon.

As we are at an early stage of the AI Act's implementation, and given the rather small amount of published summaries available, we have requested the four Providers who have provided their public summaries to discuss our findings. Through this, we aim to improve our own understanding of Provider's perspectives in completing the template, and to inform these Providers on how their summaries can be improved in quality. We consider this to be an important step as early practices in publishing the summaries might shape those that come later, and establishing a high level of quality from the onset would also be beneficial to encourage larger Providers to publish their summaries with sufficient details. Of the four Providers, only Apertus responded at the time of writing this. In our meeting, Apertus confirmed the identified issues and mentioned to us that they will take this into account in their next public summary which they are planning to be published soon. They also informed us about their perception of this work as being potentially helpful to have for small/open providers to help fill in the summary.

\subsection{Author's Perspectives and Self-Reflection}
The public summary is a powerful obligation considering its potential to mandate transparency regarding the data used to train GPAI models. The motivations of the authors in taking on this work were rooted in concerns around the understanding and enforcement of regulations as well as a desire to support the objectives of the AI Act in upholding rights and promoting responsible innovation. 
In creating the documentation quality assessment framework and carrying out the assessments, our experiences were shaped by the understanding of the intended objectives of the template, as well as practical limitations that might be faced by Providers. Therefore, in this section we share key observations to assist other evaluators, in particular the AI Office, in their assessment of public summaries.

Starting with the template, initially we found that it was sufficiently legible from a technical perspective as we were familiar with earlier efforts such as data sheets and model cards. However, when we attempted to understand which information would be `ideal' for each field and how it could be assessed across summaries, we faced barriers in achieving sufficient understanding.
This stemmed primarily from not knowing the precise rationale for why specific fields were chosen for inclusion while others which seemed conceivable were not present. In particular, we were conscious of the final template development process occurring behind closed doors, and that the discussion and choices that ultimately influenced the design and structure of the template were inaccessible to us.

Existing literature on quality assessments, both for technical documentation and data quality, places a strong emphasis on requiring structure as a way to ensure quality and ease assessments. The narrative structure of the template thus created difficulties in developing our framework and assessing it in a consistent manner as we had to take into account significant variation in the form in which documentation was provided. This complexity largely influenced how we chose metrics and combined them in dimensions to assess quality.
If the template were more structured, or provided more guidance that restricted information to expected forms, it would have vastly simplified the framework development and assessment process.
In addition, the template, despite aiming to provide a common format and structure to data summaries, was found to result in heterogenous implementations which deviated from the goal of providing the document in a standardised format. This further increased the labour involved in assessment.

Lastly, we encountered difficulties in discovering published summaries, as no defined common mechanism for publishing summaries exists, nor is there an established practice on where and how to provide the summary alongside models. This also resulted in a significant challenge as we had to expend considerable efforts to carry out an exhaustive search with the concern of missing published summaries.
These experiences motivated us to formulate recommendations for GPAI Providers to publish their public summaries in a manner that reduces friction for other stakeholders, and for the AI Office to consider improvements to the template in subsequent revisions.

\subsection{Recommendations}

\subsubsection{Recommendations for GPAI Providers} As the most prominent challenge we surfaced through this work is the lack of available public summaries, We provide the following recommendations for GPAI providers:
\begin{enumerate}[label=\small{\fbox{\tbf{P.\arabic*.}}}]
    \item Ensure the public summary document adheres closely, ideally exactly, to the template provided by the AI office.
    \item Mention the public summary document in relevant webpages and documents, for example the \textit{README} in a repository, and under the correct title as utilised by the template.
    \item For sections in the template for which no information is available or where the fields are not applicable, explicitly state this to avoid ambiguity arising from open-world assumptions.
    \item Utilise correct and high-quality SEO metadata to help surface published summaries in search engine results.
    \item Provide a common area or location for all relevant legal documentation through which stakeholders can also access the published summaries.
    \item Maintain versions of summaries and clearly indicate where to find these in the document.
    \item Ensure the public summaries contain all pertinent information, and refrain from directing stakeholders to other locations to access this information. Links to external resources that expand upon the details provided in the public summary are helpful but should not be used to avoid providing the information in the public summary.
    \item To ensure the public summaries are filled in to a high degree, having the correct and up to date information about datasets used in the model development process is necessary, which in turn requires good data governance.
\end{enumerate}

\subsubsection{Recommendations for AI Office to improve the template}:
\begin{enumerate}[label=\small{\fbox{\tbf{R.\arabic*.}}}]
    \item Adding more structured questions, i.e. non-narrative, related to provenance and metadata of the document will assist stakeholders in use of summary (e.g. regarding version, published date, link to previous versions).
    \item A broad document-level field that indicates what to expect, for example, whether the published summary relates to a new model, or a modified model; and if the base model's published summary is available (if yes, a link to it), and if not, then a justification.
    Ideally, this could be implemented through different versions of the template.
    \item A guide attached to the template that describes which information should be expected for specific fields would be helpful. Though this information is provided in part within the 'help text' part of the field itself, we found that this could be expanded to cover more use-cases and offer greater guidance.
    \item Fields as currently should be tested for use by specific stakeholders, for example through a workshop that identifies whether information is clear, comprehensible, useful, etc.
    \item Specific sections could benefit obligations for other laws beyond AI Act. For example, Section 2.4 User Data regarding GDPR Articles 13 and 14 where providers could provide information (e.g. purpose, legal basis) or a link to external resource (e.g. privacy policy).
    \item The public summaries being difficult to discover is a barrier for stakeholders. The AI Office could consider making a centralised portal that hosts all public summaries, and which would includes an online form for guiding providers in providing specific information. This would  strengthen and simplify transparency and accountability for public summaries and would simplify the enforcement of this obligation at scale.
\end{enumerate}

\subsection{Public dissemination}
Our objective of assisting GPAI Providers in filling in the template with the highest possible quality, as well as highlighting the deficiencies in published summaries, motivates us to further publish our work in a publicly visible manner.
For this, we have published our methodology, assessments, and helpful resources such as the recommendations through a public website at \url{https://aial.ie/research/gpai-training-transparency}. 
The website also acts as a common resource and provides visibility for GPAI Providers to identify other summaries and thereby improve their own, and benefits stakeholders by collecting identified public summaries in a single location (with links to their source). Our assessments provide both convenient (e.g. letter grades, percentage) and quantitative (i.e. detailed metrics) scores that are helpful to a variety of stakeholders, and the website can provide the necessary information depending on the role and interests of the stakeholder. 
In order to ensure long-term sustainability of the work, the AI Accountability Lab in Trinity College Dublin will host and oversee the publishing and maintenance of the website and associated resources, which provides a degree of sustainability and longevity to this work. 

The intended audience of the website consists of three key stakeholders: first, the GPAI model providers who can use the website to find examples of public summaries and identify best practices for implementing their own based on the quality assessments; second, the authorities and enforcers such as the AI Office for the AI Act as well as GDPR authorities, who will benefit from a common repository of public summaries and their analysis in a manner that aligns with their investigatory concerns; third, the organisations, representatives, and citizens who are concerned regarding their rights and want to understand how the public summaries could provide transparency and meaningful opportunities for them to take action. 
To allow the intended audiences to meet their objectives, we identified functional requirements regarding what information must be provided to stakeholders (e.g. metrics used in assessments, outcomes), the granularity and comprehensibility of presenting this information (e.g. overview of overall scores, followed by section-level scores, followed by a link to the full report), traceability (e.g. a link to the public summary, access to archive copies), and support (e.g. a link to AI Office guidance, relevant literature). We also identified non-functional requirements related to ease of use (e.g. design of webpages, density of information), accessibility (e.g. desktops, smartphones), and discoverability (e.g. SEO, search engines).
For maintainability and sustainability, we open open the work as a community resource via GitHub at \url{https://github.com/AIAccountabilityLab/gpai-training-transparency} where we will host our reports of all model providers in a transparent and trackable manner, and also invite contributions, discussions, and bug fixes.

\subsection{Limitations}
Our work takes place in the early stages of AI Act implementation, as is evident by the limited number of published summaries. As such, our results may not be indicative of the general practice as enforcement takes effect. However, the documentation quality assessment framework we have developed is based on robust practices that have been established within the quality assessment literature. To mitigate this, we have consciously chosen a methodological approach that permits future modifications and corrections without a large amount of repeated work to update the metrics and scores. We are thus aware and have taken into account that the developed framework will need revisions as more summaries are published and as complexities arise in understanding and using the information.

It is conceivable that our metrics and assessments may be questioned, especially by specific Providers. This can happen for many reasons, for example disagreements regarding what is of more importance, the perceived fairness of the scoring system, or even to avoid exposing problems in published documents.
As with all quality assessments, ours is one possible method for evaluating the published summaries. We have provided sufficient information to justify our approach as being informed in literature, clarified our objectives as based in pragmatic implementation of the AI Act's objectives, and identified socio-technical approaches to improve the use of the templates and published summaries through our recommendations. However, we invite future discussions regarding the implementation of our documentation quality assessment framework as we perceive the work to require a community effort in the future based on the number of GPAI Providers and published summaries to be assessed.

We also acknowledge the lack of stakeholder participation or validation as a limitation of this work. In particular, we identify the need to assess our work with Providers, stakeholders with specific rights interests such as IP, copyright, and GDPR that want to use the template, civil society organisations, and the AI Office and other enforcers. However, we consider this as an intended and planned iteration of our quality assessment framework, where the adjustments made to the metrics and scores will reflect the growing discussion and perception of transparency and usefulness within the public summaries. We have shared this work through the website and engaged with the Providers who have published their public summaries as a way to start this process.

\section{Conclusion}

The requirement for public summaries in AI Act's Article 53(1)(d) is intended to facilitate stakeholders in understanding the implications of GPAI models and to enforce their rights under EU law. As such, the quality of public summaries is a determining factor for whether stakeholders can carry out relevant investigations and take corrective actions. The documentation quality assessment framework we have developed is the first approach to consider the practicalities of this obligation and how it is implemented and assessed. Our work demonstrates both the specifics of the information demanded by the public summary, and provides valuable guidance on how the public summary and its implementations can be provided so as to achieve the stated objectives of transparency and rights empowerment.

We show how the industry is yet to react significantly to Article 53(1)(d), with only a handful of GPAI model providers having yet published their training data summaries. Of the providers which published a summary, three were small organizations with an emphasis on data transparency, including two based in EU. While major GPAI providers are yet to publish their summaries, this paper provides sufficient evidence that training data summaries can be provided in high-quality, and that the onus is on model providers to ensure that sufficient quality standards are met.
Our work also provides potentially valuable insights for the AI Office by providing a methodological tool that can show how different summaries compare in terms of transparency and usefulness. The structured analysis we developed is particularly useful for identifying whether deficiencies or issues are widespread or merely isolated to a few actors.

To assist both model providers and the AI Office, we provide actionable recommendations based on our experiences, and aim to publish training data summary assessments as a public resource. 
Finally, by publishing our assessments of public summaries \textit{publicly}, we assist stakeholders by providing all summaries in a single centralised location, and also encourage providers to provide high-quality data summaries in fulfilment of European regulations.

\section*{Acknowledgements}
\subsubsection*{Funding}
This work was funded by Mozilla.org.
\subsubsection*{Additional Funding Declarations}
The AI Accountability Lab is funded under
the AI Collaborative, an Initiative of the Omidyar Group; the Bestseller Foundation; and the John D. and Catherine T. MacArthur Foundation.
The ADAPT Research Ireland Centre for Digital Media Technology is funded by Science Foundation Ireland through the SFI Research Centres Programme and is co-funded under the European Regional Development Fund (ERDF) through Grant\#13/RC/2106\_P2. 

\subsubsection*{Generative AI Usage Statement}

The authors hereby declare that no form of generative AI was used in the writing of this paper.

\subsubsection*{Authors}
This report was authored by Dick Blankvoort and Harshvardhan J. Pandit. The authors thank Maximillian Gahntz for their role in provisioning and supervision of the project, and Zuzanna Warso, Prof. Abeba Birhane and the AI Accountability Lab in Trinity College Dublin for supporting the work.

\clearpage
\bibliographystyle{ACM-Reference-Format}
\bibliography{sample-base}
\clearpage
\appendix

\section{Detailed score cards for evaluated models}
\label{tab:scores-expanded}
\small
\begin{longtable}[]{@{}lllllllll@{}}
\caption{The detailed score cards for each of the evaluated models.}\\
\toprule\noalign{}
\tbf{SmolLM} & Clarity & Completeness & Consistency & Correctness &
Accessibility & Comprehension & Transparency & Usefulness \\
\midrule\noalign{}
\endfirsthead
\toprule\noalign{}
\tbf{SmolLM} & Clarity & Completeness & Consistency & Correctness &
Accessibility & Comprehension & Transparency & Usefulness \\
\midrule\noalign{}
\endhead
\bottomrule\noalign{}
\endlastfoot
Document & \cc{68}.18\% & \cc{100}.00\% & \cc{68}.75\% & \cc{50}.00\% & \cc{65}.38\% & \cc{50}.00\% & \cc{68}.97\% & \cc{58}.00\% \\
General information & \cc{52}.50\% & \cc{83}.08\% & \cc{100}.00\% & \cc{73}.91\% & \cc{N/A} & \cc{90}.91\% & \cc{73}.53\% & \cc{90}.91\% \\
Public Data Sources & \cc{84}.62\% & \cc{100}.00\% & \cc{100}.00\% & \cc{100}.00\% & \cc{100}.00\% & \cc{100}.00\% & \cc{96}.06\% & \cc{100}.00\% \\
Private Data Sources & \cc{N/A} & \cc{100}.00\% & \cc{100}.00\% & \cc{100}.00\% & \cc{N/A} & \cc{N/A} & \cc{100}.00\% & \cc{N/A} \\
Scraped/Crawled Data & \cc{N/A} & \cc{100}.00\% & \cc{N/A} & \cc{100}.00\% & \cc{N/A} & \cc{N/A} & \cc{100}.00\% & \cc{N/A} \\
User Data & \cc{N/A} & \cc{100}.00\% & \cc{N/A} & \cc{100}.00\% & \cc{N/A} & \cc{N/A} & \cc{100}.00\% & \cc{N/A} \\
Synthetic \& Other Data & \cc{67}.57\% & \cc{61}.29\% & \cc{100}.00\% & \cc{41}.94\% & \cc{N/A} & \cc{0}.00\% & \cc{60}.00\% & \cc{0}.00\% \\
Data Processing & \cc{100}.00\% & \cc{83}.33\% & \cc{100}.00\% & \cc{91}.67\% & \cc{100}.00\% & \cc{100}.00\% & \cc{92}.00\% & \cc{100}.00\% \\
Sum & \cc{73}.10\% & \cc{89}.12\% & \cc{95}.10\% & \cc{76}.74\% & \cc{90}.63\% & \cc{84}.17\% & \cc{82}.50\% & \cc{86}.01\% \\
& & & & & & & \tbf{B+} & \tbf{B+} \\
\end{longtable}

\begin{longtable}[]{@{}lllllllll@{}}
\toprule\noalign{}
\tbf{Apertus} & Clarity & Completeness & Consistency & Correctness &
Accessibility & Comprehension & Transparency & Usefulness \\
\midrule\noalign{}
\endhead
\bottomrule\noalign{}
\endlastfoot
Document & \cc{81}.82\% & \cc{100}.00\% & \cc{100}.00\% & \cc{70}.00\% & \cc{100}.00\% & \cc{66}.67\% & \cc{87}.04\% & \cc{84}.21\% \\
General information & \cc{47}.62\% & \cc{82}.26\% & \cc{100}.00\% & \cc{100}.00\% & \cc{100}.00\% & \cc{90}.91\% & \cc{74}.62\% & \cc{93}.75\% \\
Public Data Sources & \cc{100}.00\% & \cc{100}.00\% & \cc{100}.00\% & \cc{100}.00\% & \cc{100}.00\% & \cc{100}.00\% & \cc{100}.00\% & \cc{100}.00\% \\
Private Data Sources & \cc{N/A} & \cc{100}.00\% & \cc{100}.00\% & \cc{100}.00\% & \cc{N/A} & \cc{N/A} & \cc{100}.00\% & \cc{N/A} \\
Scraped/Crawled Data & \cc{N/A} & \cc{100}.00\% & \cc{N/A} & \cc{100}.00\% & \cc{N/A} & \cc{N/A} & \cc{100}.00\% & \cc{N/A} \\
User Data & \cc{N/A} & \cc{100}.00\% & \cc{N/A} & \cc{100}.00\% & \cc{N/A} & \cc{N/A} & \cc{100}.00\% & \cc{N/A} \\
Synthetic \& Other Data & \cc{N/A} & \cc{100}.00\% & \cc{100}.00\% & \cc{100}.00\% & \cc{N/A} & \cc{N/A} & \cc{100}.00\% & \cc{N/A} \\
Data Processing & \cc{100}.00\% & \cc{100}.00\% & \cc{100}.00\% & \cc{100}.00\% & \cc{100}.00\% & \cc{100}.00\% & \cc{100}.00\% & \cc{100}.00\% \\
Sum & \cc{82}.25\% & \cc{94}.39\% & \cc{100}.00\% & \cc{98}.48\% & \cc{100}.00\% & \cc{95}.56\% & \cc{92}.90\% & \cc{97}.14\% \\
& & & & & & & \tbf{A} & \tbf{A+} \\
\end{longtable}

\begin{longtable}[]{@{}lllllllll@{}}
\toprule\noalign{}
\tbf{Bielik} & Clarity & Completeness & Consistency & Correctness &
Accessibility & Comprehension & Transparency & Usefulness \\
\midrule\noalign{}
\endhead
\bottomrule\noalign{}
\endlastfoot
Document & \cc{81}.82\% & \cc{100}.00\% & \cc{100}.00\% & \cc{80}.00\% & \cc{100}.00\% & \cc{66}.67\% & \cc{88}.89\% & \cc{84}.21\% \\
General information & \cc{78}.89\% & \cc{86}.59\% & \cc{100}.00\% & \cc{100}.00\% & \cc{100}.00\% & \cc{90}.91\% & \cc{87}.03\% & \cc{93}.75\% \\
Public Data Sources & \cc{100}.00\% & \cc{74}.29\% & \cc{100}.00\% & \cc{100}.00\% & \cc{100}.00\% & \cc{100}.00\% & \cc{86}.91\% & \cc{100}.00\% \\
Private Data Sources & \cc{N/A} & \cc{100}.00\% & \cc{100}.00\% & \cc{100}.00\% & \cc{N/A} & \cc{N/A} & \cc{100}.00\% & \cc{N/A} \\
Scraped/Crawled Data & \cc{91}.60\% & \cc{81}.48\% & \cc{100}.00\% & \cc{77}.88\% & \cc{0}.00\% & \cc{67}.95\% & \cc{83}.78\% & \cc{51}.46\% \\
User Data & \cc{N/A} & \cc{100}.00\% & \cc{100}.00\% & \cc{100}.00\% & \cc{N/A} & \cc{N/A} & \cc{100}.00\% & \cc{N/A} \\
Synthetic \& Other Data & \cc{100}.00\% & \cc{100}.00\% & \cc{100}.00\% & \cc{100}.00\% & \cc{N/A} & \cc{100}.00\% & \cc{100}.00\% & \cc{100}.00\% \\
Data Processing & \cc{94}.74\% & \cc{100}.00\% & \cc{100}.00\% & \cc{100}.00\% & \cc{0}.00\% & \cc{86}.96\% & \cc{98}.72\% & \cc{71}.43\% \\
Sum & \cc{92}.74\% & \cc{82}.48\% & \cc{100}.00\% & \cc{88}.65\% & \cc{46}.67\% & \cc{80}.51\% & \cc{88}.02\% & \cc{71}.11\% \\
& & & & & & & \tbf{B+} & \tbf{C+} \\
\end{longtable}

\begin{longtable}[]{@{}lllllllll@{}}
\toprule\noalign{}
\tbf{Phi} & Clarity & Completeness & Consistency & Correctness & Accessibility
& Comprehension & Transparency & Usefulness \\
\midrule\noalign{}
\endhead
\bottomrule\noalign{}
\endlastfoot
Document & \cc{81}.82\% & \cc{100}.00\% & \cc{100}.00\% & \cc{70}.00\% & \cc{95}.00\% & \cc{66}.67\% & \cc{87}.04\% & \cc{81}.58\% \\
General information & \cc{50}.00\% & \cc{78}.95\% & \cc{100}.00\% & \cc{53}.85\% & \cc{N/A} & \cc{100}.00\% & \cc{70}.59\% & \cc{100}.00\% \\
Public Data Sources & \cc{0}.00\% & \cc{1}.67\% & \cc{100}.00\% & \cc{5}.00\% & \cc{0}.00\% & \cc{0}.00\% & \cc{3}.70\% & \cc{0}.00\% \\
Private Data Sources & \cc{N/A} & \cc{40}.00\% & \cc{100}.00\% & \cc{40}.00\% & \cc{N/A} & \cc{N/A} & \cc{53}.85\% & \cc{N/A} \\
Scraped/Crawled Data & \cc{N/A} & \cc{0}.00\% & \cc{N/A} & \cc{0}.00\% & \cc{N/A} & \cc{N/A} & \cc{0}.00\% & \cc{N/A} \\
User Data & \cc{N/A} & \cc{0}.00\% & \cc{N/A} & \cc{0}.00\% & \cc{N/A} & \cc{N/A} & \cc{0}.00\% & \cc{N/A} \\
Synthetic \& Other Data & \cc{0}.00\% & \cc{3}.33\% & \cc{100}.00\% & \cc{3}.33\% & \cc{N/A} & \cc{0}.00\% & \cc{8}.89\% & \cc{0}.00\% \\
Data Processing & \cc{100}.00\% & \cc{0}.00\% & \cc{100}.00\% & \cc{50}.00\% & \cc{0}.00\% & \cc{0}.00\% & \cc{52}.00\% & \cc{0}.00\% \\
Sum & \cc{28}.89\% & \cc{28}.65\% & \cc{100}.00\% & \cc{27}.08\% & \cc{21}.11\% & \cc{26}.98\% & \cc{33}.30\% & \cc{24}.54\% \\
& & & & & & & \tbf{D} & \tbf{F} \\
\end{longtable}

\begin{longtable}[]{@{}lllllllll@{}}
\toprule\noalign{}
\tbf{Bria} & Clarity & Completeness & Consistency & Correctness &
Accessibility & Comprehension & Transparency & Usefulness \\
\midrule\noalign{}
\endhead
\bottomrule\noalign{}
\endlastfoot
Document & \cc{81}.82\% & \cc{25}.00\% & \cc{100}.00\% & \cc{40}.00\% & \cc{70}.00\% & \cc{66}.67\% & \cc{70}.37\% & \cc{68}.42\% \\
General information & \cc{81}.25\% & \cc{60}.26\% & \cc{100}.00\% & \cc{71}.43\% & \cc{N/A} & \cc{92}.86\% & \cc{69}.73\% & \cc{92}.86\% \\
Public Data Sources & \cc{N/A} & \cc{100}.00\% & \cc{100}.00\% & \cc{100}.00\% & \cc{N/A} & \cc{N/A} & \cc{100}.00\% & \cc{N/A} \\
Private Data Sources & \cc{100}.00\% & \cc{100}.00\% & \cc{100}.00\% & \cc{100}.00\% & \cc{N/A} & \cc{100}.00\% & \cc{100}.00\% & \cc{100}.00\% \\
Scraped/Crawled Data & \cc{N/A} & \cc{100}.00\% & \cc{100}.00\% & \cc{100}.00\% & \cc{N/A} & \cc{N/A} & \cc{100}.00\% & \cc{N/A} \\
User Data & \cc{N/A} & \cc{100}.00\% & \cc{100}.00\% & \cc{100}.00\% & \cc{N/A} & \cc{N/A} & \cc{100}.00\% & \cc{N/A} \\
Synthetic \& Other Data & \cc{N/A} & \cc{100}.00\% & \cc{100}.00\% & \cc{100}.00\% & \cc{N/A} & \cc{N/A} & \cc{100}.00\% & \cc{N/A} \\
Data Processing & \cc{100}.00\% & \cc{100}.00\% & \cc{100}.00\% & \cc{100}.00\% & \cc{100}.00\% & \cc{100}.00\% & \cc{100}.00\% & \cc{100}.00\% \\
Sum & \cc{90}.59\% & \cc{79}.76\% & \cc{100}.00\% & \cc{90}.82\% & \cc{88}.00\% & \cc{96}.30\% & \cc{86}.97\% & \cc{94}.74\% \\
& & & & & & & \tbf{B+} & \tbf{A} \\
\end{longtable}

\clearpage
\section{List of Metrics Identified for Assessment of Public Summaries}
\label{tab:scored-assessment}
\begin{longtable}[]{@{}|p{1cm}p{8cm}p{1cm}p{1cm}p{1.75cm}P{1cm}|P{1cm}P{0.8cm}P{0.8cm}P{0.8cm}|@{}}\noalign{}
\caption{The metrics identified for template assessment, along with their weights}\\
\hline
\tbf{ID} & \tbf{Metric} & \tbf{Section} & \tbf{Field} & \tbf{Dimension} & \tbf{Weight} \\
\hline
\endfirsthead
\hline
\tbf{ID} & \tbf{Metric} & \tbf{Section} & \tbf{Field} & \tbf{Dimension} & \tbf{Weight} \\
\hline
\endhead
\hline
\endfoot
\hline
\endlastfoot
D1 & All information must be provided within a single document & 0 & n/a
& Accessibility & 3 \\
D2 & Document must be easy to find & 0 & n/a & Accessibility & 1 \\
D3 & Document must be accessible & 0 & n/a & Accessibility & 1 \\
D4 & Document should be comprehensible & 0 & n/a & Comprehension & 3 \\
D5 & Document should have the correct structure & 0 & n/a & Clarity &
1 \\
D6 & Document must be readable & 0 & n/a & Comprehension & 3 \\
D7 & Document should have clear provenance & 0 & n/a & Clarity & 1 \\
D8 & Document should have assured integrity & 0 & n/a & Correctness &
2 \\
D9 & Document should be in a well-defined, structured, and interoperable
format & 0 & n/a & Clarity & 1 \\
D10 & Document should support sharing and exporting & 0 & n/a &
Accessibility & 1 \\
D11 & Document language should be consistent with the language used in
other external documentation & 0 & n/a & Consistency & 1 \\
D12 & Document should be provided in the same context as the model & 0 &
n/a & Accessibility & 3 \\
D13 & Document should be consistent with other documents provided
elsewhere representing the same model version(s) and summary & 0 & n/a &
Consistency & 3 \\
D14 & Document should be consistent across versions (across updates) & 0
& n/a & Consistency & 3 \\
D15 & Document should clearly indicate changes from previous version & 0
& n/a & Comprehension & 2 \\
D16 & Document should clearly indicate its current status, in particular
whether it is the latest version or if it is outdated and a replacement
is made available & 0 & n/a & Clarity & 1 \\
D17 & Document should indicate where notice of updates or changes will
be provided & 0 & n/a & Comprehension & 1 \\
D18 & Document should provide link to authoritative source of the
document & 0 & n/a & Correctness & 1 \\
D19 & Document should provide link to all versions of the document & 0 &
n/a & Clarity & 1 \\
D20 & Document with updated information must be provided in the same
context as the earlier document & 0 & n/a & Clarity & 3 \\
D21 & Document should be provided in a timely manner & 0 & n/a & Clarity
& 3 \\
D22 & Document export preserves links to external resources & 0 & n/a &
Accessibility & 1 \\
& & & & & \\
& & & & & \\
& \tbf{Version of the Summary} & & & & \\
D23 & Is a version for the document provided? & 0 & 0.a & Completeness &
3 \\
D24 & Are link(s) to previous versions of the document provided, where
applicable? & 0 & 0.a & Accessibility & 3 \\
D25 & If link(s) to previous versions are provided, does each version
have a unique version number? & 0 & 0.a & Consistency & 1 \\
D26 & If each version has a unique version number, do these version
numbers adhere to a consistent format so that it is clear which version
comes prior and which one comes next? & 0 & 0.a & Correctness & 1 \\
D27 & If link(s) to previous versions are provided, are the links
accessible for intended stakeholders? & 0 & 0.a & Comprehension & 3 \\
& & & & & \\
& \tbf{Last update} & & & & \\
D28 & Does the document have a date of last update? & 0 & 0.b &
Completeness & 1 \\
D29 & Is this date accurate? & 0 & 0.b & Correctness & 1 \\
D30 & Is the date format correct? & 0 & 0.b & Correctness & 1 \\
& & & & & \\
& & & & & \\
\Large\tbf{1} & \Large\tbf{General information} & & & & \\
F1.1 & Is the information provided in this field consistent with other
sections and other documentation provided for the same model(s)? & 1 & 1
& Consistency & 3 \\
\large\tbf{1.1} & \large\tbf{Provider identification} & & & & \\
\tbf{1.1.a} & \tbf{Provider name and contact details} & & & & \\
F1.1.a.1 & Is the provider name given? & 1 & 1.1.a & Completeness & 5 \\
F1.1.a.2 & Is this provider name sufficiently detailed? (I.e. can be
traced back to individual organizations) & 1 & 1.1.a & Comprehension &
5 \\
F1.1.a.3 & Are contact details provided? & 1 & 1.1.a & Completeness &
5 \\
F1.1.a.4 & Do these contact details refer to a non-generic contact
point? (E.g. not the address used for general correspondence with the
company) & 1 & 1.1.a & Clarity & 5 \\
& & & & & \\
\tbf{1.1.b} & \tbf{Authorised representative name and contact details} & & & & \\
F1.1.b.1 & Is the name of the authorised representative given, if
required? & 1 & 1.1.b & Completeness & 3 \\
F1.1.b.2 & If the name is required, is this name sufficiently detailed?
(i.e. can be traced back to individual representatives) & 1 & 1.1.b &
Comprehension & 3 \\
F1.1.b.3 & If required, are contact details for the authorised
representative provided? & 1 & 1.1.b & Completeness & 3 \\
F1.1.b.4 & If required, do these contact details refer to a non-generic
contact point? (E.g. not the address used for general correspondence
with the company) & 1 & 1.1.b & Clarity & 3 \\
& & & & & \\
& & & & & \\
\large\tbf{1.2} & \large\tbf{Model identification} & & & & \\
\tbf{1.2.a} & \tbf{Versioned model name(s)} & & & & \\
F1.2.a.1 & Are the identifier(s) provided for each specific model(s)? &
1 & 1.2.a & Completeness & 5 \\
F1.2.a.2 & Are the provided identifiers traceable and point to a single
specific model? & 1 & 1.2.a & Comprehension & 5 \\
F1.2.a.3 & Are links to additional publicly available documentation
provided for the model(s)? & 1 & 1.2.a & Completeness & 5 \\
F1.2.a.4 & If links are provided, do these links lead to relevant
locations? & 1 & 1.2.a & Accessibility & 5 \\
F1.2.a.5 & Is the content for all provided model version(s) identical? &
1 & 1.2.a & Consistency & 5 \\
& & & & & \\
\tbf{1.2.b} & \tbf{Model dependencies} & & & & \\
F1.2.b.1 & Are dependencies indicated in cases where the model is
fine-tuned from another model? & 1 & 1.2.b & Completeness & 5 \\
F1.2.b.2 & If dependencies are indicated, is it clear for which models
they apple? & 1 & 1.2.b & Clarity & 5 \\
F1.2.b.3 & If dependencies exist, are these dependencies accurate? & 1 &
1.2.b & Correctness & 5 \\
F1.2.b.4 & If dependencies exist and there exist Summary(ies) for them,
are these linked? & 1 & 1.2.b & Completeness & 5 \\
F1.2.b.5 & If Summary(ies) are linked, do these links lead to relevant
locations? & 1 & 1.2.b & Accessibility & 5 \\
& & & & & \\
\tbf{1.2.c} & \tbf{Date of placement of the model on the Union market} & & & & \\
F1.2.c.1 & Is/are the date(s) on which the model(s) was/were placed on
the Union market given? & 1 & 1.2.c & Completeness & 1 \\
F1.2.c.2 & Are dates given accurate? & 1 & 1.2.c & Correctness & 1 \\
& & & & & \\
& & & & & \\
\large\tbf{1.3} & \large\tbf{Modalities, overall training data size and other characteristics} &
& & & \\
\tbf{1.3.a} & \tbf{Modality} & & & & \\
F1.3.a.1 & Are any modalities checked? & 1 & 1.3.a & Completeness & 3 \\
F1.3.a.2 & Are the checked modalities accurate? & 1 & 1.3.a &
Correctness & 3 \\
& & & & & \\
\tbf{1.3.b} & \tbf{Training Data Size} & & & & \\
F1.3.b.1 & For the regular modalities, is exactly only of the checkboxes
ticked, or alternatively is the approximate size provided? & 1 & 1.3.b &
Completeness & 3 \\
F1.3.b.2 & For the \textquotesingle Text\textquotesingle{} modality, if
an approximate size is provided, is the unit of measurement sensible? &
1 & 1.3.b & Clarity & 3 \\
F1.3.b.3 & For the regular modalities, if the unit of measurement is
sensible or one of the checkboxes is ticked, is the number of tokens
accurate? & 1 & 1.3.b & Correctness & 3 \\
F1.3.b.4 & For the \textquotesingle Other\textquotesingle{} modality,
are the other modalities listed sensibly? & 1 & 1.3.b & Clarity & 3 \\
F1.3.b.5 & For the \textquotesingle Other\textquotesingle{} modality,
are the units of measurement provided sensibly? & 1 & 1.3.b & Clarity &
3 \\
F1.3.b.6 & For the \textquotesingle Other\textquotesingle{} modality,
are the measurements provided accurate? & 1 & 1.3.b & Correctness & 3 \\
F1.3.b.7 & Does the \textquotesingle Other\textquotesingle{} modality
cover all modalities used in the training of the model outside of
\textquotesingle Text\textquotesingle,
\textquotesingle Image\textquotesingle,
\textquotesingle Audio\textquotesingle, and
\textquotesingle Video\textquotesingle? & 1 & 1.3.b & Completeness &
3 \\
& & & & & \\
\tbf{1.3.c} & \tbf{Types of content} & & & & \\
F1.3.c.1 & Are any types of content provided? & 1 & 1.3.c & Completeness
& 10 \\
F1.3.c.2 & If yes, does the description cover all types of content? & 1
& 1.3.c & Completeness & 10 \\
F1.3.c.3 & Does the description restrict itself to only describing the
types of content? In particular, does it refrain from describing the
goals for which the content is included? & 1 & 1.3.c & Clarity & 10 \\
& & & & & \\
\tbf{1.3.d} & \tbf{Latest date of data acquisition/collection for model training} &
& & & \\
F1.3.d.1 & Is the latest date of data collection/attainment provided? &
1 & 1.3.d & Completeness & 1 \\
F1.3.d.2 & Is this date provided exactly (i.e. exact month of latest
acquisition/collection) rather than as a loose description? & 1 & 1.3.d
& Correctness & 1 \\
F1.3.d.3 & If yes, does this date have the proper format? & 1 & 1.3.d &
Clarity & 1 \\
F1.3.d.4 & Is the date provided accurate? & 1 & 1.3.d & Correctness &
1 \\
F1.3.d.5 & Is it indicated whether the model is continuously trained on
new or dynamic data after the date provided? & 1 & 1.3.d & Comprehension
& 1 \\
F1.3.d.6 & If yes, is this information true? & 1 & 1.3.d & Correctness &
1 \\
& & & & & \\
\tbf{1.3.e} & \tbf{Description of the linguistic characteristics of the overall
training data} & & & & \\
F1.3.e.1 & Are the languages covered by the training data described? & 1
& 1.3.e & Completeness & 6 \\
F1.3.e.2 & If yes, are these languages described exactly (i.e. as a list
of languages rather than as the number covered)? & 1 & 1.3.e & Clarity &
6 \\
F1.3.e.3 & Are the EU languages which are covered mentioned? & 1 & 1.3.e
& Completeness & 6 \\
F1.3.e.4 & Is provided information relating to the linguistic
characteristics of the overall training data accurate? & 1 & 1.3.e &
Correctness & 6 \\
F1.3.e.5 & Does the field not contain any superfluous information? & 1 &
1.3.e & Clarity & 6 \\
& & & & & \\
\tbf{1.3.f} & \tbf{Other relevant characteristics of the overall training data} & &
& & \\
F1.3.f.1 & Are national/regional specificities of the training data
provided? & 1 & 1.3.f & Completeness & 6 \\
F1.3.f.2 & Are demographic specificities of the training data provided?
& 1 & 1.3.f & Completeness & 6 \\
F1.3.f.3 & Is any other relevant information relating to the
characteristics of the overall training data provided? & 1 & 1.3.f &
Clarity & 6 \\
F1.3.f.4 & Is the provided data accurate? & 1 & 1.3.f & Correctness &
6 \\
F1.3.f.5 & Is all provided information relevant? In particular, does the
field not contain any information which should be provided elsewhere in
the template? & 1 & 1.3.f & Clarity & 6 \\
& & & & & \\
\tbf{1.3.g} & \tbf{Additional comments (optional)} & & & & \\
F1.3.g.1 & Is information regarding the compression methodologies
applied for the data size calculation supplied? & 1 & 1.3.g &
Completeness & 2 \\
F1.3.g.2 & Is information regarding the tokenization methodologies
applied for the data size calculation supplied? & 1 & 1.3.g &
Completeness & 2 \\
F1.3.g.3 & If audio or video content is provided, is information
provided regarding the sampling frequency or rate plays? & 1 & 1.3.g &
Completeness & 2 \\
F1.3.g.4 & Is any other relevant information disclosed? & 1 & 1.3.g &
Completeness & 2 \\
F1.3.g.5 & Is all information contained within this field relevant? & 1
& 1.3.g & Clarity & 2 \\
& & & & & \\
& & & & & \\
\Large\tbf{2} & \Large\tbf{SECTION 2} & & & & \\
\large\tbf{2.1} & \large\tbf{Publicly available datasets} & & & & \\
F2.1.1 & Is the information provided in this field consistent with other
sections and other documentation provided for the same model(s)? & 2.1 &
2.1 & Consistency & 3 \\
\tbf{2.1.a} & \tbf{Have you used publicly available datasets to train the model?} &
& & & \\
F2.1.a.1 & Is this field filled in? & 2.1 & 2.1.a & Completeness & 1 \\
F2.1.a.2 & Is this field accurate? & 2.1 & 2.1.a & Correctness & 1 \\
& & & & & \\
\tbf{2.1.b} & \tbf{If yes, specify the modality(ies) of the content covered by the
datasets concerned} & & & & \\
F2.1.b.1 & Are any modalities supplied? & 2.1 & 2.1.b & Completeness &
3 \\
F2.1.b.2 & Are all provided modalities accurate? & 2.1 & 2.1.b &
Correctness & 3 \\
& & & & & \\
\tbf{2.1.c} & \tbf{List of large publicly available datasets} & & & & \\
F2.1.c.1 & Does this field only provide a list of large publicly
available datasets, with explanations for selecting part of the datasets
where necessary? & 2.1 & 2.1.c & Completeness & 20 \\
F2.1.c.2 & Are no large publicly available datasets which were used left
out? & 2.1 & 2.1.c & Completeness & 20 \\
F2.1.c.3 & Is a link provided for each identifier/name listed? & 2.1 &
2.1.c & Accessibility & 20 \\
F2.1.c.4 & If yes, does each identifier/name match the identifier/name
of the dataset linked? & 2.1 & 2.1.c & Consistency & 20 \\
F2.1.c.5 & Are specific versions of the datasets listed? & 2.1 & 2.1.c &
Clarity & 20 \\
F2.1.c.6 & Is a general approach to selecting part of the datasets
provided wherever part of a dataset was selected? & 2.1 & 2.1.c &
Clarity & 20 \\
F2.1.c.7 & Are the descriptions clearly understandable for the
approaches to selecting part of the datasets? & 2.1 & 2.1.c & Clarity &
20 \\
& & & & & \\
\tbf{2.1.d} & \tbf{General description of other publicly available datasets not
listed above} & & & & \\
F2.1.d.1 & Are general descriptions provided for publicly available
datasets which were used and not listed above? & 2.1 & 2.1.d &
Completeness & 16 \\
F2.1.d.2 & Does the field only contain such descriptions? & 2.1 & 2.1.d
& Clarity & 16 \\
F2.1.d.3 & Do the descriptions contain information relating to the types
of modality contained in the datasets? & 2.1 & 2.1.d & Completeness &
16 \\
F2.1.d.4 & Do the descriptions contain information relating to the
nature of the content contained in the datasets? & 2.1 & 2.1.d &
Completeness & 16 \\
F2.1.d.5 & Do the descriptions contain information relating to the
linguistic characteristics of the datasets, where applicable? & 2.1 &
2.1.d & Completeness & 16 \\
F2.1.d.6 & Do the descriptions contain information relating to the
approximate start and end dates of the data collection, or otherwise
contain that such information is ``not known"? & 2.1 & 2.1.d &
Completeness & 16 \\
F2.1.d.7 & Do the descriptions contain any other relevant information? &
2.1 & 2.1.d & Completeness & 16 \\
F2.1.d.8 & Are the descriptions provided clearly readable, such that
someone without expertise could understand them? & 2.1 & 2.1.d &
Comprehension & 16 \\
F2.1.d.9 & Do the descriptions limit themselves to relevant information?
& 2.1 & 2.1.d & Clarity & 16 \\
F2.1.d.10 & Are all descriptions accurate? & 2.1 & 2.1.d & Correctness &
16 \\
& & & & & \\
\tbf{2.1.e} & \tbf{Additional comments (optional)} & & & & \\
F2.1.e.1 & Is information relating to the size of the datasets provided?
& 2.1 & 2.1.e & Completeness & 3 \\
F2.1.e.2 & Are other relevant details provided in this field? & 2.1 &
2.1.e & Completeness & 3 \\
F2.1.e.3 & Is all information provided in this field relevant? & 2.1 &
2.1.e & Comprehension & 3 \\
F2.1.e.4 & Is all information provided in this field accurate? & 2.1 &
2.1.e & Correctness & 3 \\
& & & & & \\
& & & & & \\
\large\tbf{2.2} & \large\tbf{Private non-publicly available datasets obtained from third
parties} & & & & \\
F2.2.1 & Is the information provided in this field consistent with other
sections and other documentation provided for the same model(s)? & 2.2 &
2.2 & Consistency & 3 \\
\large\tbf{2.2.1} & \large\tbf{Datasets commercially licensed by rightsholders or their
representatives} & & & & \\
\tbf{2.2.1.a} & \tbf{Have you concluded transactional commercial licensing
agreement(s) with rightholder(s) or with their representatives?} & & &
& \\
F2.2.1.a.1 & Is this field filled in? & 2.2 & 2.2.1.a & Completeness &
1 \\
F2.2.1.a.2 & Is this field accurate? & 2.2 & 2.2.1.a & Correctness &
1 \\
& & & & & \\
\tbf{2.2.1.b} & \tbf{If yes, specify the modality(ies) of the content covered by
the datasets concerned} & & & & \\
F2.2.1.b.1 & Are any modalities supplied? & 2.2 & 2.2.1.b & Completeness
& 3 \\
F2.2.1.b.2 & Are all provided modalities accurate? & 2.2 & 2.2.1.b &
Correctness & 3 \\
& & & & & \\
\large\tbf{2.2.2} & \large\tbf{Private datasets obtained from other third parties} & & & & \\
\tbf{2.2.2.a} & \tbf{Have you obtained private datasets from third parties that
are not licensed as described in Section 2.2.1., such as data obtained
from providers of private databases, or data intermediaries?} & & & & \\
F2.2.2.a.1 & Is this field filled in? & 2.2 & 2.2.2.a & Completeness &
1 \\
F2.2.2.a.2 & Is this field accurate? & 2.2 & 2.2.2.a & Correctness &
1 \\
& & & & & \\
\tbf{2.2.2.b} & \tbf{If yes, specify the modality(ies) of the content covered by
the datasets concerned} & & & & \\
F2.2.2.b.1 & Are any modalities supplied? & 2.2 & 2.2.2.b & Completeness
& 3 \\
F2.2.2.b.2 & Are all provided modalities accurate? & 2.2 & 2.2.2.b &
Correctness & 3 \\
& & & & & \\
\tbf{2.2.2.c} & \tbf{If publicly known, list private datasets obtained from other
third parties} & & & & \\
F2.2.2.c.1 & Does this field only provide a list of publicly known
private datasets obtained from other third parties without license? &
2.2 & 2.2.2.c & Clarity & 20 \\
F2.2.2.c.2 & Are any publicly known private datasets obtained from third
parties without license, which were used for this model, left out? & 2.2
& 2.2.2.c & Correctness & 20 \\
F2.2.2.c.3 & Are specific versions of the datasets listed? & 2.2 &
2.2.2.c & Clarity & 20 \\
F2.2.2.c.4 & Are links to relevant information provided? & 2.2 & 2.2.2.c
& Accessibility & 20 \\
& & & & & \\
\tbf{2.2.2.d} & \tbf{General description of non-publicly known private datasets
obtained from third parties} & & & & \\
F2.2.2.d.1 & Are general descriptions provided for all non-publicly
known, non-licensed private datasets obtained from third parties which
were used? & 2.2 & 2.2.2.d & Completeness & 16 \\
F2.2.2.d.2 & Does the field only contain such descriptions? & 2.2 &
2.2.2.d & Clarity & 16 \\
F2.2.2.d.3 & Do the descriptions contain information relating to the
types of modality contained in the datasets? & 2.2 & 2.2.2.d &
Completeness & 16 \\
F2.2.2.d.4 & Do the descriptions contain information relating to the
nature of the content contained in the datasets? & 2.2 & 2.2.2.d &
Completeness & 16 \\
F2.2.2.d.5 & Do the descriptions contain information relating to the
linguistic characteristics of the datasets, where applicable? & 2.2 &
2.2.2.d & Completeness & 16 \\
F2.2.2.d.6 & Do the descriptions contain any other relevant information?
& 2.2 & 2.2.2.d & Completeness & 16 \\
F2.2.2.d.7 & Are the descriptions provided clearly readable, such that
someone without expertise could understand them? & 2.2 & 2.2.2.d &
Comprehension & 16 \\
F2.2.2.d.8 & Do the descriptions limit themselves to relevant
information? & 2.2 & 2.2.2.d & Clarity & 16 \\
F2.2.2.d.9 & Are all descriptions accurate? & 2.2 & 2.2.2.d &
Correctness & 16 \\
& & & & & \\
\tbf{2.2.2.e} & \tbf{Additional comments (optional)} & & & & \\
F2.2.2.e.1 & Is information relating to the period of data collection
provided? & 2.2 & 2.2.2.e & Completeness & 3 \\
F2.2.2.e.2 & Is information relating to the size of the datasets
provided? & 2.2 & 2.2.2.e & Completeness & 3 \\
F2.2.2.e.3 & Are other relevant details provided in this field? & 2.2 &
2.2.2.e & Completeness & 3 \\
F2.2.2.e.4 & Is all information provided in this field relevant? & 2.2 &
2.2.2.e & Clarity & 3 \\
F2.2.2.e.5 & Is all information provided in this field accurate? & 2.2 &
2.2.2.e & Correctness & 3 \\
& & & & & \\
& & & & & \\
\large\tbf{2.3} & \large\tbf{Data crawled and scraped from online sources} & & & & \\
F2.3.1 & Is the information provided in this field consistent with other
sections and other documentation provided for the same model(s)? & 2.3 &
2.3 & Consistency & 3 \\
\tbf{2.3.a} & \tbf{Were crawlers used by the provider or on behalf of?} & & & & \\
F2.3.a.1 & Is this field filled in? & 2.3 & 2.3.a & Completeness & 3 \\
F2.3.a.2 & Is this field accurate? & 2.3 & 2.3.a & Correctness & 3 \\
& & & & & \\
\tbf{2.3.b} & \tbf{If yes, specify crawler name(s)/identifier(s)} & & & & \\
F2.3.b.1 & Does the field only contain crawler name(s)/identifier(s)? &
2.3 & 2.3.b & Clarity & 15 \\
F2.3.b.2 & Can these name(s)/identifier(s) each be used to identify
specific crawlers? & 2.3 & 2.3.b & Clarity & 15 \\
F2.3.b.3 & Were all crawlers which are listed indeed used? & 2.3 & 2.3.b
& Correctness & 15 \\
F2.3.b.4 & Are all crawlers which were used described? & 2.3 & 2.3.b &
Completeness & 15 \\
F2.3.b.5 & Is the information provided in a clearly-structured form? &
2.3 & 2.3.b & Clarity & 15 \\
& & & & & \\
\tbf{2.3.c} & \tbf{Purposes of the crawler(s)} & & & & \\
F2.3.c.1 & Are descriptions relating to the purposes of the crawler(s)
provided? & 2.3 & 2.3.c & Completeness & 10 \\
F2.3.c.2 & Are descriptions provided for each crawler? & 2.3 & 2.3.c &
Completeness & 10 \\
F2.3.c.3 & Are these descriptions understandable, such that someone
without specialised expertise can understand the purposes of the
crawlers? & 2.3 & 2.3.c & Comprehension & 10 \\
F2.3.c.4 & Does the field only contain information relating to the
purposes of the crawlers listed? & 2.3 & 2.3.c & Clarity & 10 \\
F2.3.c.5 & Are all of the descriptions accurate? & 2.3 & 2.3.c &
Correctness & 10 \\
& & & & & \\
\tbf{2.3.d} & \tbf{General description of crawler behaviour} & & & & \\
F2.3.d.1 & Is a general description of crawler behavior provided? & 2.3
& 2.3.d & Completeness & 10 \\
F2.3.d.2 & If a description is provided, does it contain information
relating to the respect of captchas by crawlers? & 2.3 & 2.3.d & Clarity
& 10 \\
F2.3.d.3 & If a description is provided, does it contain information
relating to the handling of password protected websites by crawlers? &
2.3 & 2.3.d & Clarity & 10 \\
F2.3.d.4 & If a description is provided, does it contain information
relating to the handling of paywalls by crawlers? & 2.3 & 2.3.d &
Clarity & 10 \\
F2.3.d.5 & If a description is provided, does it contain information
relating to the respect of robots.txt by crawlers? & 2.3 & 2.3.d &
Clarity & 10 \\
F2.3.d.6 & If a description is provided, does it contain information
relating to the respect of protocols outside of robots.txt by crawlers?
& 2.3 & 2.3.d & Clarity & 10 \\
F2.3.d.7 & Does the description contain other relevant information? &
2.3 & 2.3.d & Completeness & 10 \\
F2.3.d.8 & Is the description clearly understandable, such that someone
without detailed technical knowledge can understand the crawler
behavior? & 2.3 & 2.3.d & Comprehension & 10 \\
F2.3.d.9 & Does the description limit itself to relevant information? &
2.3 & 2.3.d & Clarity & 10 \\
F2.3.d.10 & Is the description accurate? & 2.3 & 2.3.d & Correctness &
10 \\
& & & & & \\
\tbf{2.3.e} & \tbf{Period of data collection} & & & & \\
F2.3.e.1 & Is a period of data collection provided? & 2.3 & 2.3.e &
Completeness & 7 \\
F2.3.e.2 & If a period of data collection is provided, is it provided
for each crawler? & 2.3 & 2.3.e & Clarity & 7 \\
F2.3.e.3 & If a period of data collection is provided, are exact dates
given? & 2.3 & 2.3.e & Clarity & 7 \\
F2.3.e.4 & If exact dates are given, are these in the appropriate format
(MM/YYYY)? & 2.3 & 2.3.e & Correctness & 7 \\
F2.3.e.5 & Is all information provided accurate? & 2.3 & 2.3.e &
Correctness & 7 \\
& & & & & \\
\tbf{2.3.f} & \tbf{Comprehensive description of the type of content and online
sources crawled} & & & & \\
F2.3.f.1 & Is information relating to the geographical characteristics
of the crawled content provided? & 2.3 & 2.3.f & Completeness & 13 \\
F2.3.f.2 & Is information relating to the linguistic characteristics of
the crawled content provided? & 2.3 & 2.3.f & Completeness & 13 \\
F2.3.f.3 & Is information relating to the demographic characteristics of
the crawled content provided? & 2.3 & 2.3.f & Completeness & 13 \\
F2.3.f.4 & Is an indication given regarding which type(s) of websites
are scraped? & 2.3 & 2.3.f & Completeness & 13 \\
F2.3.f.5 & Is any other pertinent information relating to the types of
content crawled provided? & 2.3 & 2.3.f & Completeness & 13 \\
F2.3.f.6 & Is all provided information relevant? & 2.3 & 2.3.f &
Comprehension & 13 \\
F2.3.f.7 & Is all provided information accurate? & 2.3 & 2.3.f &
Correctness & 13 \\
& & & & & \\
\tbf{2.3.g} & \tbf{Type of modality covered} & & & & \\
F2.3.g.1 & Are any modalities supplied? & 2.3 & 2.3.g & Completeness &
3 \\
F2.3.g.2 & Are all provided modalities accurate? & 2.3 & 2.3.g &
Correctness & 3 \\
& & & & & \\
\tbf{2.3.h} & \tbf{Summary of the most relevant domain names crawled} & & & & \\
F2.3.h.1 & Is a list of the most relevant internet domains provided as
per the requirements? & 2.3 & 2.3.h & Completeness & 25 \\
F2.3.h.2 & Is the provided list accurate? & 2.3 & 2.3.h & Correctness &
25 \\
F2.3.h.3 & Is the provided list easily accessible? & 2.3 & 2.3.h &
Accessibility & 25 \\
F2.3.h.4 & Is the provided list provided in a straightforwardly readable
format? & 2.3 & 2.3.h & Comprehension & 25 \\
& & & & & \\
\tbf{2.3.i} & \tbf{Additional comments (optional)} & & & & \\
F2.3.i.1 & In this field, are more domains disclosed than those required
in the list above? & 2.3 & 2.3.i & Clarity & 20 \\
F2.3.i.2 & Are the URLs and sources of individual works provided in this
field? & 2.3 & 2.3.i & Clarity & 20 \\
F2.3.i.3 & Are other relevant details provided in this field? & 2.3 &
2.3.i & Clarity & 20 \\
F2.3.i.4 & Is all information provided in this field relevant? & 2.3 &
2.3.i & Comprehension & 20 \\
F2.3.i.5 & Is all information provided in this field accurate? & 2.3 &
2.3.i & Correctness & 20 \\
& & & & & \\
& & & & & \\
\large\tbf{2.4} & \large\tbf{User data} & & & & \\
F2.4.1 & Is the information provided in this field consistent with other
sections and other documentation provided for the same model(s)? & 2.4 &
2.4 & Consistency & 3 \\
\tbf{2.4.a} & \tbf{Was data from user interactions with the AI model (e.g. user
input and prompts) used to train the model?} & & & & \\
F2.4.a.1 & Is this field filled in? & 2.4 & 2.4.a & Completeness & 3 \\
F2.4.a.2 & Is this field accurate? & 2.4 & 2.4.a & Correctness & 3 \\
& & & & & \\
\tbf{2.4.b} & \tbf{Was data collected from user interactions with the
provider\textquotesingle s other services or products used to train the
model?} & & & & \\
F2.4.b.1 & Is this field filled in? & 2.4 & 2.4.b & Completeness & 5 \\
F2.4.b.2 & Is this field accurate? & 2.4 & 2.4.b & Correctness & 5 \\
& & & & & \\
\tbf{2.4.c} & \tbf{If yes, provide a general description of the
provider\textquotesingle s services or products that were used to
collect the user data} & & & & \\
F2.4.c.1 & Does this field only contain the relevant services or
products by the provider used to train the model? & 2.4 & 2.4.c &
Clarity & 15 \\
F2.4.c.2 & Are the services or products each clearly identified? & 2.4 &
2.4.c & Comprehension & 15 \\
F2.4.c.3 & Are all involved services or products by the provider listed?
& 2.4 & 2.4.c & Completeness & 15 \\
F2.4.c.4 & Is the description provided in a readable format? & 2.4 &
2.4.c & Accessibility & 15 \\
& & & & & \\
\tbf{2.4.d} & \tbf{Type of modality covered} & & & & \\
F2.4.d.1 & Are any modalities supplied? & 2.4 & 2.4.d & Completeness &
3 \\
F2.4.d.2 & Are all provided modalities accurate? & 2.4 & 2.4.d &
Correctness & 3 \\
& & & & & \\
\tbf{2.4.e} & \tbf{Additional comments (optional)} & & & & \\
F2.4.e.1 & Are relevant details provided in this field? & 2.4 & 2.4.e &
Completeness & 1 \\
F2.4.e.2 & Is all information provided in this field relevant? & 2.4 &
2.4.e & Clarity & 1 \\
F2.4.e.3 & Is all information provided in this field accurate? & 2.4 &
2.4.e & Correctness & 1 \\
& & & & & \\
& & & & & \\
\large\tbf{2.5} & \large\tbf{Synthetic data} & & & & \\
F2.5.1 & Is the information provided in this field consistent with other
sections and other documentation provided for the same model(s)? & 2.5 &
2.5 & Consistency & 3 \\
\tbf{2.5.a} & \tbf{Was synthetic AI-generated data created by the provider or on
their behalf to train the model?} & & & & \\
F2.5.a.1 & Is this field filled in? & 2.5 & 2.5.a & Completeness & 1 \\
F2.5.a.2 & Is this field accurate? & 2.5 & 2.5.a & Correctness & 1 \\
& & & & & \\
\tbf{2.5.b} & \tbf{If yes, modality of the synthetic data} & & & & \\
F2.5.b.1 & Are any modalities supplied? & 2.5 & 2.5.b & Completeness &
3 \\
F2.5.b.2 & Are all provided modalities accurate? & 2.5 & 2.5.b &
Correctness & 3 \\
& & & & & \\
\tbf{2.5.c} & \tbf{If yes, specify the general-purpose AI model(s) used to generate
the synthetic data if available on the market} & & & & \\
F2.5.c.1 & If yes was ticked, are models provided? & 2.5 & 2.5.c &
Completeness & 12 \\
F2.5.c.2 & If models are provided, are these uniquely identifiable? &
2.5 & 2.5.c & Clarity & 12 \\
F2.5.c.3 & Are the correct models provided? & 2.5 & 2.5.c & Correctness
& 12 \\
F2.5.c.4 & Is a link to the Summary(ies) of the relevant models provided
where such a summary is available? & 2.5 & 2.5.c & Comprehension & 12 \\
F2.5.c.5 & Does the field only contain the list of GPAI models used and
a link to their Summary(ies) where available? & 2.5 & 2.5.c & Clarity &
12 \\
& & & & & \\
\tbf{2.5.d} & \tbf{Information about other AI models, includer
provider\textquotesingle s own AI model(s) not available on the market,
used to generate synthetic data to train the model to which this Summary
applies} & & & & \\
F2.5.d.1 & Is information provided for each AI model used to generate
synthetic data which is not available on the market? & 2.5 & 2.5.d &
Completeness & 12 \\
F2.5.d.2 & Does the field only contain information about the relevant
group of AI models? & 2.5 & 2.5.d & Clarity & 12 \\
F2.5.d.3 & Does the information provided include a general description
of each model\textquotesingle s training data if known and necessary as
described? & 2.5 & 2.5.d & Comprehension & 12 \\
F2.5.d.4 & Is any other relevant information about the relevant group of
AI models provided? & 2.5 & 2.5.d & Clarity & 12 \\
F2.5.d.5 & Is all provided information accurate? & 2.5 & 2.5.d &
Correctness & 12 \\
& & & & & \\
\tbf{2.5.e} & \tbf{Additional comments (optional)} & & & & \\
F2.5.e.1 & Is relevant information provided in this field? & 2.5 & 2.5.e
& Completeness & 1 \\
F2.5.e.2 & Is all information provided in this field relevant? & 2.5 &
2.5.e & Clarity & 1 \\
F2.5.e.3 & Is all information provided in this field accurate? & 2.5 &
2.5.e & Correctness & 1 \\
& & & & & \\
& & & & & \\
\large\tbf{2.6} & \large\tbf{Other sources of data} & & & & \\
F2.6.1 & Is the information provided in this field consistent with other
sections and other documentation provided for the same model(s)? & 2.5 &
2.6 & Consistency & 3 \\
\tbf{2.6.a} & \tbf{Have data sources other than those described in Sections 2.1 to
2.5. been used to train the model?} & & & & \\
F2.6.a.1 & Is this field filled in? & 2.5 & 2.6.a & Completeness & 2 \\
F2.6.a.2 & Is this field accurate? & 2.5 & 2.6.a & Correctness & 2 \\
& & & & & \\
\tbf{2.6.b} & \tbf{If yes, provide a narrative description of these data sources
and the data} & & & & \\
F2.6.b.1 & Does this field contain relevant information relating to the
miscellaneous data sources and their data? & 2.5 & 2.6.b & Completeness
& 17 \\
F2.6.b.2 & Does this field restrict itself to relevant information
relating to the miscellaneous data sources and their data? & 2.5 & 2.6.b
& Clarity & 17 \\
F2.6.b.3 & Is all information accurate? & 2.5 & 2.6.b & Correctness &
17 \\
F2.6.b.4 & Is the information provided in a clearly readable format? &
2.5 & 2.6.b & Accessibility & 17 \\
& & & & & \\
\tbf{2.6.c} & \tbf{Additional comments (optional), for other sources of data} & & &
& \\
F2.6.c.1 & Is relevant information provided in this field? & 2.5 & 2.6.c
& Completeness & 1 \\
F2.6.c.2 & Is all information provided in this field relevant? & 2.5 &
2.6.c & Correctness & 1 \\
F2.6.c.3 & Is all information provided in this field accurate? & 2.5 &
2.6.c & Accessibility & 1 \\
& & & & & \\
& & & & & \\
\Large\tbf{3} & \Large\tbf{SECTION 3} & & & & \\
F3.1 & Is the information provided in this field consistent with other
sections and other documentation provided for the same model(s)? & 3 & 3
& Consistency & 3 \\
\large\tbf{3.1} & \large\tbf{Respect of reservation of rights from text and data mining
exception or limitation} & & & & \\
\tbf{3.1.a} & \tbf{Are you a Signatory to the Code of Practice for general-purpose
AI models that includes commitments to respect reservations of rights
from the TDM exception or limitation?} & & & & \\
F3.1.a.1 & Is this field filled in? & 3 & 3.1.a & Completeness & 3 \\
F3.1.a.2 & Is this field accurate? & 3 & 3.1.a & Correctness & 3 \\
& & & & & \\
\tbf{3.1.b} & \tbf{Describe the measures implemented before model training to
respect reservations of rights from the TDM exception or limitation
before and during data collection, including the opt-out proticols and
solutions honoured by the provider or, as applicable, by third parties
from which datasets have been obtained} & & & & \\
F3.1.b.1 & Does the provider describe how they respect robots.txt? & 3 &
3.1.b & Comprehension & 18 \\
F3.1.b.2 & Does the provider describe how they use datasets with opt-out
mechanisms, if applicable? & 3 & 3.1.b & Comprehension & 18 \\
F3.1.b.3 & Does the provider describe how they ensure that they are up
to date with user rights requests? & 3 & 3.1.b & Comprehension & 18 \\
F3.1.b.4 & Is all other information relevant? & 3 & 3.1.b & Clarity &
18 \\
F3.1.b.5 & Is all information provided accurate? & 3 & 3.1.b &
Correctness & 18 \\
& & & & & \\
\tbf{3.1.c} & \tbf{Additional comments (optional)} & & & & \\
F3.1.c.1 & Is a summary of the providers copyright policy provided, if
this policy is made publicly available? & 3 & 3.1.c & Completeness &
5 \\
F3.1.c.2 & Is all other information provided relevant? & 3 & 3.1.c &
Clarity & 5 \\
& & & & & \\
& & & & & \\
\large\tbf{3.2} & \large\tbf{Removal of illegal content} & & & & \\
\tbf{3.2.a} & \tbf{General decription of measures taken} & & & & \\
F3.2.a.1 & Are any measures described to avoid or remove illegal content
under Union law from the training data? & 3 & 3.2.a & Completeness &
15 \\
F3.2.a.2 & If measures are described, are these measures described in an
easily understandable manner? & 3 & 3.2.a & Comprehension & 15 \\
F3.2.a.3 & If measures are described, is the information provided
accurate? & 3 & 3.2.a & Correctness & 15 \\
F3.2.a.4 & Is all information provided pertinent? In particular, does
the information not include an information relating to data selection
practices, for example to increase the capability of the model? & 3 &
3.2.a & Accessibility & 15 \\
& & & & & \\
& & & & & \\
\large\tbf{3.3} & \large\tbf{Other information (optional)} & & & & \\
\tbf{3.3.a} & \tbf{Other relevant information about data processing (optional)} & &
& & \\
F3.3.a.1 & Is any additional information provided? & 3 & 3.3.a &
Completeness & 1 \\
F3.3.a.2 & If yes, is this relevant to data processing aspects and
measures taken before or after training model that his relevant for the
respect and exercise of rights protected under Union law? & 3 & 3.3.a &
Clarity & 1 \\
F3.3.a.3 & If yes, is this information accurate? & 3 & 3.3.a &
Correctness & 1 \\
\end{longtable}

\end{document}